\documentclass[apj]{emulateapj}
\submitted{Submitted to Astrophysical Journal}

\shorttitle{AGN Broad-Line Region Geometry} \shortauthors{Gaskell,
Klimek, \& Nazarova}

\begin{document}

\title{NGC 5548: THE AGN ENERGY BUDGET PROBLEM AND THE GEOMETRY \\
OF THE BROAD-LINE REGION AND TORUS}

\author{C. MARTIN GASKELL\altaffilmark{1} AND ELIZABETH S. KLIMEK\altaffilmark{2}}
\affil{Department of Physics \& Astronomy, University of Nebraska,
Lincoln, NE 68588-0111}

\and

\author{LUDMILA S. NAZAROVA\altaffilmark{3,4}}

\affil{Euro Asian Astronomical Society, Universitetskij Pr. 13,
Moscow 119899, Russia}

\altaffiltext{1}{Present Address: Department of Astronomy,
University of Texas, Austin, TX 78712-0259. Electronic address:
gaskell@astro.as.utexas.edu}

\altaffiltext{2}{Present Address: Department of Astronomy, New
Mexico State University, Las Cruces, NM 88003-0001.  Electronic
address: eklimek@nmsu.edu}

\altaffiltext{3}{Electronic address: lsn@kzn.ru}

\altaffiltext{4}{Visiting scientist, Department of Physics and
Astronomy, University of Nebraska}

\begin{abstract}

We consider in detail the spectral energy distribution (SED) and
multi-wavelength variability of NGC~5548.  Comparison with the SEDs
of other AGNs implies that the internal reddening of NGC~5548 is
E(B-V) = 0.17 mag. The extinction curve is consistent with the mean
curve of other AGNs found by Gaskell \& Benker, but inconsistent
with an SMC-type reddening curve.  Because most IR emission
originates exterior to the broad-line region (BLR), the SED seen by
the inner BLR is different from that seen by the outer BLR and from
the earth. The most likely BLR covering factor is $\sim 40$\% and it
is not possible to get an overall BLR covering factor of less than
20\%. This requires that the BLR is not spherically symmetric and
that we are viewing through a hole. Line-continuum variability
transfer functions are consistent with this geometry. The covering
factor and geometry imply that near the equatorial plane the BLR
covering approaches 100\%. The spectrum seen by the outer regions of
the BLR and by the torus is thus modified by the absorption in the
inner BLR. This shielding solves the problem of observed BLR
ionization stratification being much greater than implied by
photoionization models.  The BLR obscuration also removes the
problem of the torus covering factor being greater than the BLR
covering factor, and gives consistency with the observed fraction of
obscured AGNs. The flux reduction at the torus also reduces the
problem of AGN dust-reverberation lags giving sizes smaller than the
dust-sublimation radii.

\end{abstract}


\section{INTRODUCTION}

Because of the difficulty of observing in the extreme ultra-violet
(EUV), the true shape of the continuum of AGNs has long been a
mystery.  If emission lines arise from photoionization, then the
equivalent width of the lines gives information on the number of
ionizing photons (Zanstra 1931, Osterbrock 1989) and their total
energy (Stoy 1933).  Early photoionization model calculations of the
broad-line region (BLR) spectrum (e.g., Davidson 1972, MacAlpine
1972, Shields 1972) {\it assumed} covering factors ($\Omega/4\pi$)
and continuum shapes that could account for the observed strengths
of the emission lines. However, the first spectrophotometry of the
rest-frame UV continuum in high-redshift AGNs showed an apparent
turn down of the UV spectrum (Oke 1970, 1974). {\it IUE} satellite
observations of lower-redshift AGNs confirmed this (Green et al.\@
1980).  Green et al.\@ reported that the $\alpha \thickapprox 0.5$
spectral slope ($F_{\nu} \propto \nu^{-\alpha}$) from the optical to
the UV steepened to $\alpha \thickapprox 2.3$ at wavelengths shorter
than $\lambda$1200. MacAlpine (1981) showed that there was a severe
``energy budget'' problem because such a continuum needed
$\Omega/4\pi \gtrsim 9$ to match the observed He II $\lambda$4686
equivalent width.  The observed ionizing continuum was thus
providing an order-of-magnitude fewer ionizing photons than were
needed.

The problem was compounded when spectroscopy of the Lyman-limit
region showed that rest-frame Lyman continuum absorption was rare
(Osmer 1979). Smith et al.\@ (1981) concluded from observations of
the Lyman-limit region that $\Omega/4\pi \lesssim 0.15$ and further
analysis by MacAlpine (1981) showed that $\Omega/4\pi \thickapprox
0.05$ was most likely.  Netzer (1985) pointed out that there is a
serious energy-budget discrepancy for other lines.  The HST
composite spectrum of Zheng et al.\@ (1997) and the multi-wavelength
spectra of a sample of PG quasars (Laor 1997) confirmed the
steepening of the observed continuum at shorter wavelengths.
Photoionization models (e.g., Shields \& Ferland 1993; Korista et
al.\@ 1998; Goad \& Koratkar 1998; Kaspi \& Netzer 1999; Korista \&
Goad 2000) consistently required covering factors an order of
magnitude higher than allowed by the limits on Lyman limit
absorption.

An important question is whether the continuum {\it we} see is the
same as that seen by the emission lines.  The continuum could be
intrinsically anisotropic, or the continuum we see could be modified
by extinction from dust grains along the line of sight, or by
absorption lines. MacAlpine (1981) suggested that a marked upturn in
dust extinction below $\lambda$1200 could be causing the apparent
turndown in the continuum we see.

Green et al.\@ (1980), however, felt that the turndown below
$\lambda$1200 was probably intrinsic, and that extinction by dust
grains was probably not causing it because of the lack of effect on
the spectrum longwards of $\lambda$1200. McKee \& Petrosian (1974)
had previously argued against the presence of dust in AGNs because
of the lack of both the $\lambda$2175 dust feature and the curvature
in the UV spectra that dust should produce.  The apparent match up
between UV spectral indices and the X-ray region (Laor 1997) argued
against significant extinction.

Various other solutions have been offered to the energy-budget
problem. Collin-Souffrin (1986), Joly (1987), and Dumont,
Collin-Souffrin, \& Nazarova (1998) favored non-radiative heating of
the BLR as the solution to the energy-balance problem. Binette et
al.\@ (1993) suggested the clouds were being heated by a diffuse
continuum very close to them. Korista, Ferland, \& Baldwin (1997)
suggested the the UV-EUV SED might be double peaked. Maiolino et
al.\@ (2001c), on the other hand, suggested that the solution to the
conflict between the lack of Lyman limit absorption and the high
covering factors implied by photoionization models was that the BLR
was not covering the source uniformly.

In this paper we investigate the energy-budget question in the
well-studied AGN NGC~5548 and investigate the implications for the
BLR and torus structure. NGC~5548 is an important test case for
investigating the energy-budget issue (Rokaki et al.\@ 1994; Dumont
et al.\@ 1998) since NGC~5548 has been particularly well studied at
all accessible wavelengths, its variability behavior in most
spectral regions is well known, and the line-continuum transfer
functions, which give information on the BLR size and radial
distribution, are better known than for any other AGN. We focus in
this paper on the period 2449115 -- 2449130 when NGC 5548 was
intensely monitored by the {\it Hubble Space Telescope}. We adopt a
distance, $R$, of 76 Mpc to NGC~5548. This is based on an H$_0 = 73$
km/s/Mpc, $\Omega_{matter} = 0.27$, $\Omega_{vacuum} = 0.73$
cosmology and the local velocity field model of Mould et al.\@
(2000).

\section{THE SPECTRAL ENERGY DISTRIBUTION OF NGC 5548}

Many different spectral energy distributions (SEDs) have been
considered when modelling NGC~5548.  A number of these SEDs are
illustrated in Fig.\@ 2 of Dumont et al.\@ (1998), in Chiang \&
Blaes (2003), and Fig.\@ 1 of Steenbrugge et al.\@ (2005). It can be
seen that the SEDs differ by up to factors of three in the region
below the Lyman limit and elsewhere. We therefore start off by
reconstructing the probable SED of NGC~5548 around JD 2449120 in
detail.

Like all AGNs, NGC~5548 varies.  It was in fact the very first AGN
for which variability was discovered (see historical discussion in
Gaskell \& Klimek 2003).  This variability has to be allowed for in
constructing the continuum shape.  As far as possible, one needs
simultaneous observations of the continuum at different wavelengths.
However, such observations are not available for all wavelength
regions, so, where there were no simultaneous observations, we have
estimated the fluxes by scaling observations from other epochs. For
determining the appropriate scalings we have used other sets of
quasi-simultaneous data.  These scalings are necessary for
estimating the SED at a given epoch, but they should also be useful
for modeling spectral variability at different epochs.

From the 73-day period JD 2449061 to 2449134 NGC~5548 was
simultaneously monitored with the {\it Hubble Space Telescope}, the
{\it IUE} satellite, and many optical observatories (Korista et
al.\@ 1995). From the period of {\it HST} coverage we have selected
JD 2449115 - 2449130 when NGC~5548 was in a high state. During this
period data were available only for some of the optical and UV
continua and emission lines, so in the following subsections we
describe, in order of construction, how the continuum in each
waveband was obtained.

\subsection{Optical and UV}

In Table 1 we give our measured optical continua from the {\it
International AGN Watch (IAW)} optical spectra (see Wanders \&
Peterson 1996\footnote{Note that all the line and continuum fluxes
in their paper are too low by a factor of two.}). Three relatively
line-free sections of the spectra between major emission lines were
taken to represent the continuum. These sections were defined by the
observed wavelength ranges: 4640--4660 \AA, 4810--4830 \AA, and
5190--5210 \AA. We measured average UV continuum points from the HST
spectra (Korista et al.\@ 1995) during the high state at
$\lambda$$\lambda$ 1170, 1360, 1490, 1820. 2050, and 2250 \AA. We
used observed wavelength ranges of 1160--180 \AA, 1344--1374 \AA,
1478-1501 \AA, 1804-1834 \AA, 2022-2204 \AA, and 2225-2241 \AA.




\begin{deluxetable}{lcrcc}
\tablewidth{0pt} \tablecaption{Observed and Corrected Continuum
Fluxes} \tablehead{ \colhead{Region} & Log($\nu$)& \colhead{Log($\nu
F_{\nu}$)\tablenotemark{a}} & \colhead{MW de-red} &
\colhead{Internal de-red}} \startdata
100 $\mu$m& 12.48 & -10.14 & & \\
60 $\mu$m& 12.70 & -10.29 & & \\
25 $\mu$m& 13.08 & -10.00 & & \\
12 $\mu$m& 13.40 & -10.03 & & \\
10.2 $\mu$m& 13.46 & -10.15 & & \\
5 $\mu$m& 13.80 & -10.32 & & \\
3.4 $\mu$m& 13.93 & -10.08 & & \\
2.2 $\mu$m& 14.13 & -10.10 & & \\
1.65 $\mu$m& 14.27 & -10.03 & & \\
1.25 $\mu$m& 14.38 & -10.06 & & \\
5200 \AA& 14.76 & -10.24 & -10.21 & -10.13 \\
4820 \AA& 14.79 & -10.25 & -10.22 & -10.10 \\
4650 \AA& 14.81 & -10.27 & -10.23 & -10.10 \\
2250 \AA& 15.12 & -10.29 & -10.20  & -9.72 \\
2050 \AA& 15.17 & -10.29 & -10.20 & -9.69 \\
1820 \AA& 15.22 & -10.24 & -10.17 & -9.63 \\
1490 \AA& 15.30 & -10.24 & -10.16 & -9.59 \\
1360 \AA& 15.34 & -10.21 & -10.13 & -9.54 \\
1179 \AA& 15.41 & -10.23 & -10.12 & -9.53 \\

\enddata
\tablenotetext{a}{$\nu F_{\nu}$ in ergs s$^{-1}$ cm$^{-2}$}
\end{deluxetable}

To obtain fluxes in other passbands, when there were no simultaneous
observations, we scaled observations to our optical or UV fluxes.
There are many broad-band optical photometric observations available
but these include host-galaxy contamination. We need to correct for
different host galaxy contributions and differing passbands. The
{\it IAW} database gives monochromatic fluxes at 5100\AA
~standardized to a $5 \times 7.6$ arcsecond spectroscopic aperture.
Romanishin et al.\@ (1995) discuss how to estimate the nuclear flux
of NGC ~5548 from direct images and they give the transformation
between the 5100\AA ~fluxes and V-band fluxes. They give estimates
of AGN-free galaxy magnitudes as a function of photometric aperture
and give the host galaxy light contribution to the standard {\it
IAW}  spectroscopic aperture. They estimate the galaxy flux in the
standard photometric aperture to be $2.53 \times 10^{-15}$ ergs
s$^{-1}$ cm$^{-2}$ \AA$^{-1}$, while the galaxy contribution in an
8-arcsecond radius photometric aperture is $5.81 \times 10^{-15}$
ergs s$^{-1}$ cm$^{-2}$ \AA$^{-1}$.

Although for our modeling below we consider the host-galaxy-free AGN
optical fluxes, for comparison with other wavelengths, our observed
fluxes (column 2 of Table 1) still include the galaxy contribution
since this is included by other observers. In section 4 we subtract
host galaxy contributions from all wavebands.


\subsection {Infra-red}


Even though infra-red radiation is not relevant in providing
ionizing photons, it does contribute to the heating of the broad
line region and hence to the total energy budget.  As we will
discuss below (see section 2.3), in order to know the IR flux
received by the BLR clouds one needs to know the distance of the IR
emitting regions from the center. It will also be important to know
the inner radius of the dusty torus because, as proposed by Netzer
\& Laor (1993) and Laor (2007), we will find that this is probably
the outer edge of the BLR.

To our knowledge, there were no simultaneous infrared observations
made at the time of the HST observations. We can, however, estimate
the IR flux from the relationship between the IR and optical at
other times. From a study of 41 Seyfert galaxies, Glass (2004)
reported that all but two showed variability, with K-band (2.2
$\mu$m) amplitudes in the range $< 0.1$ to $>1.1$ mag. The shortest
timescale he found for detectable changes was one week. Glass found
that, in general, the near-IR flux varies in the same way as the
optical, UV, and X-ray flux, but with a time delay. For NGC~5548 he
reports observations in the J, H, K, \& L bands made on 12 nights
over the three-year period 1988 - 1990. Lyutyi \& Doroshenko (1993)
give UBV photometry during the same period.

    The delay of the IR fluxes relative to the optical and UV is
substantial so the IR flux depends on the optical flux at an earlier
time. Furthermore there will be an additional delay in IR radiation
from the dust getting back to the BLR clouds.  The lag of the IR
flux was not measured for NGC 5548 in the high state we consider,
but we can estimate the lag by scaling from the measurements of the
lag by Suganuma et al.\@ (2004) when NGC 5548 was in a low state and
independently by scaling from other objects.


Glass (2004) gives estimated delays and $3.4 \mu$m luminosities for
17 AGNs. In Fig.\@ 1 we show the relationship between the IR lag,
$\tau$, and luminosity. A least squares fit gives

\begin{equation}
\log(\tau) = 0.48 \log(L_{3.4\mu m}) -9.14
\end{equation}

This correlation is completely consistent with the expected $\tau
\propto L^{0.5}$ relationship for a constant dust sublimation
temperature.  Minezaki et al.\@ (2005) and Suganuma et al.\@ (2006)
find a similar correlation between $\tau$ and the V-band absolute
magnitude.  For NGC~5548, $L_{3.4\mu m} = 22.81$, so Fig.\@ 1
implies a delay of $\sim 84$ days.

\begin{figure} \plotone{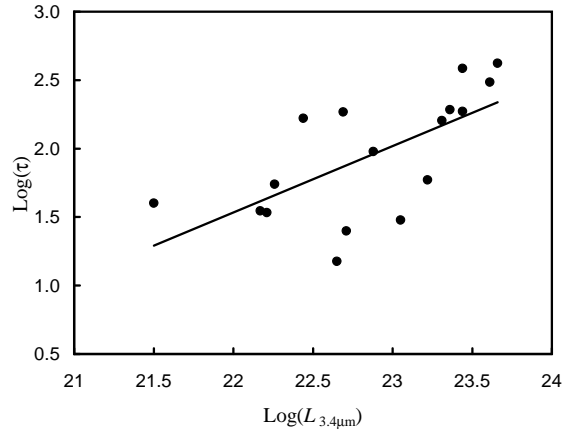} \caption{IR lags,
$\tau$ (in days), as a function of L-band flux (log WHz$^{-1}$)
for 17 AGNs. The lags are the responsivity-weighted lags of the J
and L band fluxes relative to the U band fluxes. The line is a
least-square fit. Data from Glass (2004).}
\end{figure}

Suganuma et al.\@ (2004) monitored NGC~5548 in the IR and optical
bands from JD 2451950 to JD 2452825, when NGC 5548 was in a low
state. They measured time delays between the K band and the V band
of $48 \pm 3$d and $47\pm 6$d. Even though the mean V-band flux of
NGC~5548 decreased by a factor of about 2.8 between the earlier
epoch we are considering, their lags can be used to check our
estimate if we consider the change in the level of activity. Due to
dust sublimation, an increase in luminosity leads to an increase in
the effective torus radius, thereby increasing the lag between the
IR and the optical according to the relationship $\tau \propto \surd
L$ (Oknyanskij et al.\@ 1999.) Thus, when the measured lag is scaled
by the ratio of the average luminosities during the two epochs, we
should get a lag consistent with our estimate. Scaling of the
Sugnuma et al.\@ lag gives 79 d, which is in good agreement with our
earlier estimate of 84 d from the general IR radius--luminosity
relationship, and within the error bars of the Suganuma et al.\@
lags.

By comparing the J-band (1.25 $\mu$m) fluxes of Glass (2004) with
the average V magnitudes 80 days earlier, from Lyutyi \&
Doroshenko (1993) we get the approximate relationship:

\begin{equation}
J = 0.4 V + 6.2
\end{equation}

To estimate the IR continuum seen {\it by the BLR} from the optical
continuum we need to look at the optical continuum, not at a time
$\tau$ earlier, but at a time $\sim 2\tau$ earlier. This is because
the UV/optical continuum heating the dust takes $\sim \tau$ days to
get to the dust torus, and the reprossed IR takes {\it another}
$\sim \tau$ days on average to get back to the BLR. Because of the
larger size of the IR-emitting region, we took the average optical
flux (from Korista\ et al.\@ 1995) between $2\tau - 50$ days before
and $2\tau + 50$ days before our high and low states. After
appropriate scaling of the Korista et al.\@ fluxes to the magnitudes
of Lyutyi \& Doroshenko (1993), we obtained J magnitudes of 11.63
and 11.66 for our high and low states. As would be expected, the IR
flux does not follow the more rapid UV/optical changes.


Choloniewski (1981) showed that, in the optical, the shape of the
{\it variable} component of AGNs remains the same as the flux level
changes (see also Winkler et al.\@ 1992). The analysis of Glass
(2004) shows that, for any given AGN, the colors of the variable
component of the IR emission are also usually independent of the
level of activity and, for his whole sample, he finds that the
colors of the variable components fall within moderately narrow
ranges. We have therefore used our calculated J magnitude and the
average IR colors of Glass to get estimated H, K, and L magnitudes
of 10.77, 10.05, and 8.76 respectively for the low state, and 10.74,
10.02, and 8.73 for the high state.  These fluxes fall well within
the historical ranges of fluxes given in the literature.


For longer IR wavelengths we have taken measurements from the
literature. Potential problems with these measurements include
inhomogeneity due to varying data quality and photometric apertures
used, and source variability. We excluded measurements that were
taken using apertures larger than 15'' diameter, as these
measurements contain significant additional off-nuclear infrared
emission from the host galaxy. For the M (4.8 $\mu$m) waveband we
averaged the fluxes of McAlary, McLaren, \& Crabtree (1979) and
McAlary et al.\@ (1983), and for the N (10 $\mu$m) waveband we
averaged the fluxes of Kleinmann \& Low (1970), Rieke \& Low (1972),
and Rieke (1978). There is observational evidence that for these
longer IR wavelengths non-simultaneity is not a problem. Neugebauer
et al.\@ (1989) found that while AGNs show variability in J, H, K,
and L, there was no variability at 10 $\mu$m. Similarly, Edelson \&
Malkan (1987) found that there was no evidence for variability in
the {\it IRAS} passbands for non-blazar AGNs.

The 12, 25, 60, and 100 $\mu$m IRAS observations (Moshir et al.\@
1990) have a low spatial resolution and no variability information
is available. They refer to a region much further out in the galaxy
than the near-IR torus.

\subsection{IR Continuum Seen by the BLR}

A large fraction of the IR continuum in an AGN is clearly not coming
from the central source, but from a torus well outside the BLR.  The
ratio of IR to optical/UV that a distant observer sees is {\it not}
the same as what a BLR cloud sees.

Calculating the X-ray, UV, and optical flux at the BLR clouds is
straight forward.  We assume that the region emitting this flux is
smaller than the BLR and located inside it.  We can then obtain the
flux by simply multiplying the flux observed at the earth by
$(R/r)^2$ where $R = 9.05 \times 10^{10}$ ld is the distance to
NGC~5548, and $r$ is the distance of the BLR clouds from the center.
For the IR, for which a substantial part of the emission observed
from the earth arises from emission from the torus {\it outside} the
BLR, the assumption of a compact source within the BLR is incorrect,
and we will overestimate the flux at the BLR if we just scale the
flux in the same way as the X-ray, UV, and optical.  Instead, the IR
flux at the BLR clouds depends on the surface brightness of the IR
emitting regions and the solid angles they subtend.

We roughly estimated the relative dilution of the IR continuum at
the BLR clouds with the following scheme.  We divided the radiation
received by a BLR cloud at radius $r_c$ into two components: the
radiation originating from  $r < r_c$ and the radiation originating
from $r \geq r_c$.  We assumed that the flux of radiation
originating at $r < r_c$ could be scaled from the flux received at
the earth by $(R/r)^2$, where $R$ is the distance to NGC~5548. If we
consider a uniform thin spherical emitting shell of radius
$r_{shell}$, the radiation density falls off as $r^{-2}$ for $r >
r_{shell}$ while it is constant for $r \leq r_{shell}$ (i.e., within
the shell). This still remains a reasonable approximation if the
emitting matter is in a ring rather than a shell. Therefore, for
radiation originating outside $r_c$ the flux was scaled by
$(R/r_{shell})^2$. Thus the IR radiation contribution from outside
the BLR was reduced by $(r_c/R_{shell})^2$ relative to the
contribution from inside the BLR.

To estimate the fraction of IR radiation from within $r_c$ we
assumed there was a minimum accretion-disk temperature of 5000 K.
Any IR flux less than that produced by the spectrum of a 5000 K
black body was assumed to come from outside $r_c$ and reduced by
$(r_c/R{shell})^2$. In a standard thin accretion disk (Pringle \&
Rees 1972, Shakura \& Sunyaev 1973) the temperature, $T$ falls off
with radius as $r^{-0.75}$. If there is reprocessing of radiation
from the inner regions to the outer regions then the temperature
gradient is shallower (e.g., $T \propto R^{-0.50}$ for the ``slim
disk'' of Abramowicz et al.\@ 1988).  If dust is in thermal
equilibrium with radiation from a compact source, $T_{dust} \propto
R^{-0.50}$ as well, so we assume that $T \propto R^{-0.50}$.  This
implies that the effective radius emitting a given wavelength is
$R_{eff} \propto \lambda^2$.  The reduction of the IR radiation is
thus $\propto \lambda^4$.  The contribution from outside the BLR can
thus be ignored at all but the shortest wavelengths.

Because of this effect, BLR clouds at different radii receive
different long-wavelength SEDs.  The outer BLR is close to the inner
edge of the torus and experiences a K-band flux similar to that
given by a simple inverse-square-law scaling; the inner BLR is an
order of magnitude closer and receives 100 times less relative
K-band flux from the dust than would be expected by simple scaling.
In our photoionization modeling we therefore use different SEDs at
different radii.

\subsection{Hard X-rays}

\subsubsection {Relationship Between Hard X-ray and Optical/UV
Flux Levels}

There were no simultaneous hard X-ray observations during the HST
campaign, so we need to estimate the hard X-ray flux from the
optical and UV. Uttley et al.\@ (2003) show that the 2--10 keV flux
measured with {\it RXTE} is well correlated on {\it long} timescales
(months) with the optical flux at $\lambda$5100. However, on short
timescales (day-to-day), the X-rays in NGC~5548 can vary a lot while
the optical varies little (see Uttley et al.\@ 2003 and Gaskell et
al.\@ in preparation). We compared the RXTE 2--10 keV fluxes from
Uttley et al.\@ (2003) with the 5100\AA~ fluxes (from the {\it IAW}
database) measured within $\pm 7$ d.  The median separation time
between the X-ray and optical observations is 0.6 d.   As can be
seen in Fig.\@ 2, there is a direct proportionality, after
subtracting the galaxy contribution as estimated by Romanishin et
al.\@ (1995). We found that, apart from short-term events on the
timescale of a month or two, the average X-ray/optical ratio was
constant in time and showed no correlation with the optical flux
level. The X-ray flux, in units of $10^{-11}$ ergs s$^{-1}$
cm$^{-2}$, can be predicted from the optical by the equation

\begin{equation}
F_{2-10 keV}= 0.564 \times F_{\lambda 5100},
\end{equation} where the $F_{\lambda 5100}$ is in units of $10^{-15}$ ergs s$^{-1}$
cm$^{-2}$ \AA$^{-1}$.

\begin{figure} \plotone{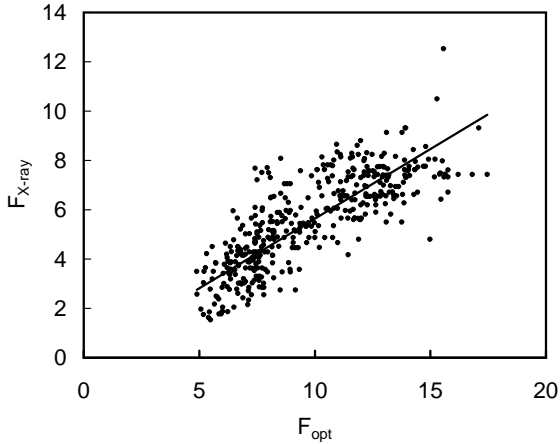} \caption{The relationship between
the 2--10keV flux (in units of $10^{-11}$ ergs s$^{-1}$ cm$^{-2}$)
and the host-galaxy subtracted optical flux (in units of $10^{-15}$
ergs s$^{-1}$ cm$^{-2}$ \AA$^{-1}$) for near simultaneous
observations. The straight line is the relationship $F_{2-10 keV}=
0.564 \times F_{\lambda 5100}$}
\end{figure}

Clavel et al.\@ (1992) (see their Fig.\@ 4) found that the 2--10~keV
flux observed with {\it Ginga}, was also directly proportional to
the 1350\AA~flux:

\begin{equation}
F_{2-10 keV} = 0.114 \times F_{\lambda 1350}.
\end{equation}

In our high state, $F_{\lambda 5100} = 6.49 \times 10^{-15}$ after
host-galaxy subtraction, and $F_{\lambda 1350} = 44.8 \times
10^{-15}$. From equations (3) and (4) these predict $F_{2-10 keV} =
3.7 \times 10^{-11}$ and $5.1 \times 10^{-11}$ ergs s$^{-1}$
cm$^{-2}$ respectively. We adopt a mean of $4.4 \times 10^{-11}$
ergs s$^{-1}$ cm$^{-2}$.

\subsubsection{Shape of the Hard X-ray Spectrum}

As in most AGNs, the shape of the hard X-ray spectrum of NGC~5548
can be described by a power law ($F_{\nu} \propto \nu^{-\alpha}$)
with a spectral index $\alpha \sim 0.7$.  In NGC 5548 $\alpha$
depends on the flux level.  Markowitz, Edelson, \& Vaughan (2003)
show, from {\it RXTE} data, that for NGC~5548 the degree of hard
X-ray variability is constant as a function of energy on short
timescales but decreases for higher energies on long timescales.
Chiang et al.\@ (2000) and Papadakis et al.\@ (2002) find a good
correlation between $\alpha$ and $F_{2-10keV}$. From Fig.\@ 4 of
Chiang et al.\@ and our estimated $F_{2-10keV}$ fluxes we get
$\alpha$ = 0.76. This is within the range found by Magdziarz et
al.\@ (1998) and also consistent with the trend found by Walter \&
Courvoisier (1990).

The hard cut-off in the $\gamma$-ray region is well constrained by
{\it OSSE} spectra from the {\it Compton Gamma Ray Observatory}.
Magdziarz et al.\@ (1998) get a hard energy cutoff of 120~keV for
1991 and 1993 {\it OSSE} observations.


\subsection{Extreme Ultraviolet and Soft X-ray Region}

The most energetically important region of the spectrum of AGNs is
the extreme ultraviolet (EUV) and soft X-ray spectral region. This
is also the region responsible for most of the ionization of the
BLR. Unfortunately, it is very hard to observe AGNs in the EUV and
soft X-ray spectral region because of heavy atomic absorption.

Walter et al.\@ (1994) present {\it ROSAT} soft X-ray observations
of several AGNs, including NGC 5548. For all objects they find a
strong soft X-ray excess above the extrapolation of the higher
energy X-rays to lower energies. They find that the soft X-ray
spectral shape is very similar in the different sources even though
they range over four orders of magnitude in luminosity. The soft
X-ray excess can be described by a bump with a similar cutoff energy
of $kT \sim 200$ keV for sources with a wide range of luminosity.
For NGC 5548 itself, the effective temperature of the soft excess
appears to remain constant at $kT \sim 400$ eV (see Fig.\@ 7b of
Magdziarz et al.\@ 1998). More recent $XMM$-$Newton$ and
$Beppo$-$SAX$ observations (Pounds et al.\@ 2003) also show a clear
soft excess below 700 eV which, after allowance for complex
absorption, probably smoothly extends to 2 keV in agreement with the
Walter et al.\@ (1994) spectrum. There is debate over whether the
derived temperature of the soft excess in AGNs is real, or whether
the apparent exponential cutoff is due instead to strong
relativistically smearing of a partially-ionized absorption
(Gierlinski \& Done 2004), but the fundamental nature of the soft
excess does not affect the continuum the broad-line region sees, so
long as the BLR sees the same continuum we see.

There was no simultaneous monitoring of the soft X-ray spectrum
during the {\it HST}/{\it IUE} campaign. We must therefore try to
estimate the EUV/soft-X-ray flux from either the UV or the hard
X-ray components. To do this we need to consider the nature of EUV
variability and the relationship of EUV and soft X-ray variability
to variability of the continuum at higher and lower energies.

\subsubsection{Variability of the Soft Component}

Our knowledge of variability of the EUV/soft X-ray flux of NGC~5548
has come primarily from monitoring with the {\it ROSAT} satellite,
covering 0.1 -- 2.4 keV, and the {\it EUVE} satellite, giving
effectively measurements at a single energy of $\sim 0.2$~keV.
Walter \& Courvoisier (1990) found that as NGC~5548 varied, although
the ratio of the soft excess to the higher energy power law
component changed, the shape of the {\it soft} X-ray spectrum did
not.  {\it EUVE} count rates cannot be readily converted into flux
units because the conversion depends on the spectral shape and
hydrogen column density (see Marshall et al.\@ 1995), but given the
apparent constancy in the spectral shape, they are important for
showing the {\it relative} variability at $\sim 0.2$~keV.

Observations reveal two important things about the soft excess. The
first is that {\it the soft excess varies most}. The ratio of
changes in the hard and soft X-ray components during the {\it ROSAT}
monitoring of Nandra et al.\@ (1993) is consistent with the soft
excess varying more, and the {\it ROSAT} monitoring of Done et al.\@
(1995) shows that variability is primarily at energies of less than
0.4 keV (see their Fig.\@ 2). From subsequent simultaneous {\it
EUVE} and X-ray monitoring of NGC~5548, Chiang et al.\@ (2000) found
that the amplitude of the {\it EUVE} variability was 15 -- 20\%
greater than that of the {\it RXTE} variability.

The second important thing is that, on short timescales, {\it the
soft excess leads the variability}.  Kaastra \& Barr (1987)
discovered from {\it EXOSAT} monitoring of NGC~5548 that the soft
X-rays led the hard X-rays by about 1.5 hours. Walter \& Courvoisier
(1990) found a similar leading of about 1 hour. Chiang et al.\@
(2000) found that the EUV (0.2~keV) led the 0.5--1 keV ASCA band in
NGC~5548 by $0.15 \pm 0.05$d, and led the RXTE 2--20 keV band by
$0.45 \pm 0.11$d. These small delays are unimportant for our
analysis, but the sign of the delay together with the greater
amplitude of variability of the soft excess implies that {\it
short-term X-ray variability is driven by variability of the soft
excess}. The width at zero intensity of the EUV/hard X-ray cross
correlation function of Chiang et al.\@ (2000) is $\pm 1$d. The size
of the EUV emitting region is thus $\lesssim 1$ lt.d.  The {\it
EUVE} light curve of Chiang et al.\@ (2000) shows there are
sometimes changes of up to a factor of two on a timescale of an
hour.  We can thus be confident that the EUV to soft X-ray emission
originates well inside the BLR.

\subsubsection{Correlation of Soft Excess with Lower Energies}

Although there were no simultaneous soft X-ray spectral measurements
during the 1993 {\it HST} campaign, Marshall et al.\@ (1997) made
{\it EUVE} observations which overlap some of the UV observations,
and despite the problem of converting {\it EUVE} counts to fluxes,
the {\it amplitude} of EUV/soft X-ray variability can be estimated
from the {\it EUVE} observations since the {\it EUVE} samples part
of the {\it ROSAT} band at $\sim 0.2$ keV. This monitoring is
important for showing the {\it relative} change in the flux of the
soft excess as a function of the UV flux.

The {\it EUVE} observations over the period JD 2409104 -- 2409112
(shortly after the high state we are considering) overlap the {\it
HST} monitoring, and other {\it EUVE} observations overlap the {\it
IUE} monitoring.  The amplitude of the EUV variability is about
$\sim 5$ times greater than the UV variability at $\lambda$1350. The
EUV light curve shows continual variations of a factor of $\sim 2$
on timescales of one or two days.  The later {\it EUVE} light curve
of Chiang et al.\@ (2000) shows similar variations. For the most
prominent event seen both in the EUV and UV spectra, Marshall et
al.\@ find a lag consistent with zero delay between the two bands.

In Fig.\@ 3 we show the correlation between the {\it EUVE} fluxes
reported by Marshall et al.\@ and the $\lambda$1350 UV flux measured
by the {\it IUE} and {\it HST}.  We have averaged {\it EUVE} fluxes
within $\pm 1$d of the UV flux measurement.   From a least-squares
fit we find that

\begin{equation}
f_{EUV} = 0.11 f_{1350} - 97.
\end{equation}

\begin{figure} \plotone{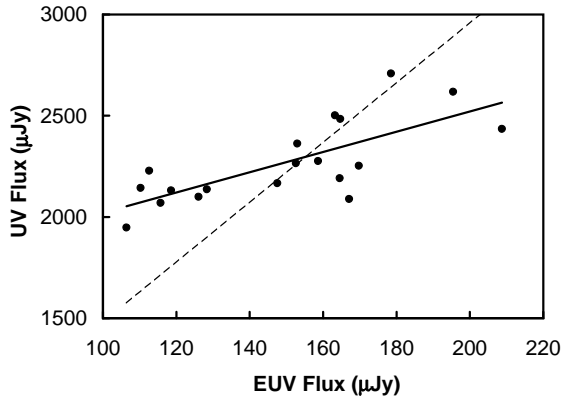} \caption{UV fluxes at 1350\AA,
observed with the {\it IUE} satellite versus {\it EUVE} 84\AA
 ~fluxes (from Marshall et al.\@ 1997) averaged over $\pm 1$d.  The
solid line is a least-squares fit. The dashed line represents a
constant ratio of EUV to UV flux.}
\end{figure}

The range of EUV variability during the Marshall et al.\@ (1997)
monitoring was a factor of two, while the UV range was only a factor
of 40\%. Equation (5) and Fig.\@ 3 imply that there is another
component of UV flux that will remain even when the EUV flux goes to
zero, at least on the timescale of the Marshall et al.\@
observations.

Support for a correlation of the EUV with longer wavelengths is also
provided by Dietrich et al.\@ (2001), who made optical observations
simultaneous with the Chiang et al.\@ (2000) EUV and hard X-ray
observations. Over the 30-day monitoring period, the optical and EUV
fluxes are correlated (see Fig.\@ 4 of Dietrich et al.\@), but the
EUV flux varied by a factor of $\gtrsim 6$ while the observed
optical flux variations were only 20\%. Dietrich et al.\@ find
optical-EUV lags consistent with zero, but with fairly large errors
of $\pm$ 2 -- 3$^d$.

We used the UV flux rather than the optical flux to estimate the
what the observed EUV/soft X-ray flux would have been during the
period we are considering because the difference in amplitude
between the UV and EUV is not as great as between the optical and
the EUV, and because the UV is closer to the EUV in log($\nu$) than
the optical is, and thus the UV has a higher probability of being
physically connected to the EUV. The estimated observed EUV flux is
not the true EUV flux because of the atomic absorption (see
Steenbrugge et al.\@ 2005 and references therein). We can get a
better estimate of the true EUV and soft X-ray flux by scaling to
the higher energy side of this region, since the absorption problems
are less. We therefore used the Walter et al.\@ (1994) soft X-ray
spectral shape relative to the comparatively absorption-free harder
X-ray flux, which has been estimated above.

\subsubsection{Relationship Between the Soft and Hard X-ray
Components}

The relationship between the soft and hard X-ray components provides
an additional consistency check on our estimate of the soft
component. As noted above, the hard X-ray variations lag the
variations of the soft excess by a fraction of a day. From
simultaneous {\it ASCA} and {\it EUVE} monitoring in 1996, Haba et
al.\@ (2003) found that, from day to day, the EUV and 2--10~keV
X-ray fluxes were well correlated. The EUVE amplitude over a 10-day
period was only 40\% greater than the hard X-ray amplitude. As noted
already, the 1998 simultaneous ASCA and EUVE monitoring by Chiang et
al.\@ (2000) showed a good correlation between the EUV and 2--10~keV
X-ray fluxes, with the EUV amplitude being only 15 -- 20\% greater
than the X-rays.  Haba et al.\@ did find, however, that on a shorter
timescale (6 hours), there was a strong 40\% dip in the EUV that had
no counterpart in the hard X-rays (see their Fig.\@ 2), but such
rapid variability is not relevant for the emission lines we study
here.

Walter \& Courvoisier (1990) fit the lowest energy observed with the
thin Lexan filter on {\it EXOSAT} with a power law extrapolated from
the hard X-rays and a soft-component.  The soft-excess generally
makes up 90\% of the flux in the filter.  In Fig.\@ 4 we show the
relationship between the soft-excess and the hard X-ray component.
There is a similar relationship between the hard X-rays and the soft
excess as between the UV and the soft excess (see Fig.\@ 3). Given
that the 2--10 keV flux is correlated with the optical and UV on the
BLR light-crossing timescale (see section 2.4), the amount of EUV
variability implied by the  relationship between the hard and soft
X-ray fluxes is similar to that implied by the optical and UV.

\begin{figure}
\plotone{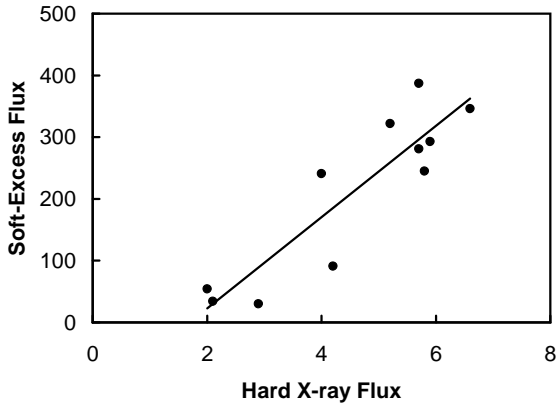} \caption{Soft Excess (in counts) vs. Hard X-ray
component (in units of $10^{-11}$ ergs s$^{-1}$ cm$^{-2}$)}
\end{figure}

\section{The Galactic Reddening Correction}

The observed continuum fluxes must first be corrected for Galactic
extinction. Schlegel et al.\@ (1998) used measurements of dust
emission at 100 and 240 $\mu$m to create Galactic extinction maps,
which have an accuracy of 16\%. From these the extinction, E(B-V),
is $0.020 \pm 0.003$ mag for the direction of NGC 5548. From 21-cm
observations Murphy et al.\@ (1996) have determined the neutral
hydrogen column density, $n_H$, towards NGC~5548 to be $1.62 \times
10^{20}$.  If we adopt a standard dust-to-gas ratio of $E(B-V) =
1.72 \times 10^{-22} n_H$ (Bohlin, Savage \& Drake 1978), we get
E(B-V) = 0.028. If adopt a minimal dust-to-gas ratio of $E(B-V) =
1.2 \times 10^{-22} n_H$ (see Fig.\@ 1 of Congiu et al.\@ 2005) we
get E(B-V) = 0.020.  We can thus be confident that the Galactic
reddening is relatively low and we adopt $E(B-V) = 0.024 \pm 0.004$

The Galactic extinction curve is well-determined for the near-IR,
optical, and UV.  For the far and extreme UV and soft X-ray regions
the shape of the extinction curve is not observationally determined
so we used the interstellar dust model of Draine (2003b) for $R =
A_V/E(B-V) = 3.1$, which is a good fit to longer wavelengths (Draine
2003a).

\section{Host Galaxy Correction}


The observed spectral energy distribution we have compiled is not
the spectral energy distribution of the AGN alone: emission from the
host galaxy is also included. Schmitt et al.\@ (1997) have made a
compilation of spectral energy distributions for normal galaxies,
which enables us to subtract off the host galaxy contributions at
various wavelengths. 
Romanishin et al.\@ (1995) find that the galaxy component of NGC
5548 is a normal early-type spiral, with no conclusive evidence of
star formation in the bulge.  We have therefore adopted the spiral
galaxy SED of Schmitt et al.\@ (1997). We scaled the spiral galaxy
SED to a flux at 5200 \AA ~of $3.6 \times 10^{-15}$ ergs s$^{-1}$
cm$^{-2}$ \AA$^{-1}$ = $3.3 \times 10^{-26}$ ergs s$^{-1}$ cm$^{-2}$
Hz$^{-1}$ in the standard spectroscopic aperture of the {\it IAW}.

\section{Internal Reddening Correction}

Gaskell \& Benker (2007) have obtained extinction curves for
individual AGNs by assuming that the unreddened continua are the
same.  For most of the AGNs they consider this gives extinction
curves which resemble the reddening curve in the local solar
neighborhood minus the $\lambda$2175 amorphous carbon feature, but
which are flatter in the UV (see also Czerny et al.\@ 2004 and
Gaskell et al.\@ 2004). For one of the AGNs they consider they find
a steeply-rising, SMC-type reddening curve.  We obtained the
extinction curve for NGC~5548 in the same manner as Gaskell \&
Benker by comparing our observed spectrum of NGC~5548 with the
observed blue spectrum of the relatively unreddened AGN 3C~273.  As
can be seen in Fig.\@ 5, the resulting reddening curve is consistent
with the mean AGN reddening curve of Gaskell \& Benker (2007) but is
inconsistent with an SMC reddening curve. Because of the consistency
with the standard AGN reddening curve we have adopted the latter for
NGC~5548.  The normalization of NGC~5548 to the standard AGN curve
gives $E(B-V) = 0.17$ mag.  This has a formal uncertainty of $\sim
\pm 0.01$ mag., but the real uncertainty is probably several times
this because of the uncertainty in the host galaxy correction for
3C~273.

\begin{figure}
\plotone{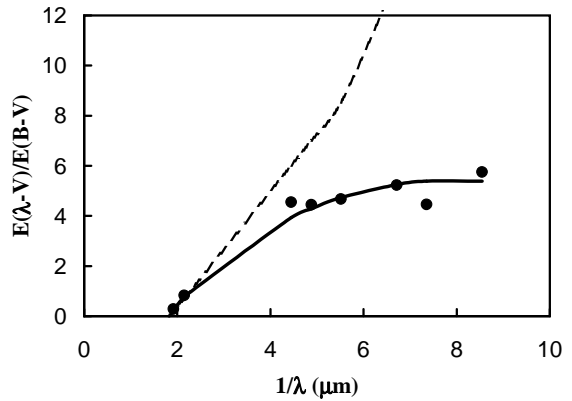} \caption{The extinction curve for NGC 5548 (filled
circles) compared with the mean AGN reddening curve of Gaskell \&
Benker (2007) (solid curve) and an SMC reddening curve (dashed
line).}
\end{figure}

If we assume the standard Galactic dust-to-gas ratio of Bohlin et
al.\@ (1978) within NGC 5548, E(B-V) = 0.17 corresponds to a total
hydrogen column of $10^{21}$ cm$^{-2}$.  This is the same as the
median column density Walter \& Courvoisier (1990) obtain.
Steenbrugge et al.\@ (2005) get total best-fit column densities for
their warm absorber components of $3 \times 10^{21}$ cm$^{-2}$ which
would imply E(B-V) $\sim 0.5$ with a standard gas/dust ratio but
this is not necessarily inconsistent with our lower reddening since
not all warm-absorber gas will have dust associated with it.

The Gaskell \& Benker (2007) reddening curve only extends to Lyman
$\alpha$.  They find that the slope of AGN spectra in the far UV
($\lambda < $ Ly $\alpha$) is not well correlated with E(B-V)
determined from UV-optical spectra (see their Fig.\@ 7) and they
propose that the scatter in the far-UV slope is the result of a
small amount of additional reddening (E(B-V) $\sim 0.03$ mag.) by a
SMC-like dust since this spectral region is very sensitive to such
dust. The bluest AGNs have a spectral index from $\sim 1200$\AA ~to
$\sim 800$\AA ~of $\sim 0.75$. We have adopted this as the
unreddened spectra slope in this spectral region.

The mean AGN extinction curve is uncertain at high energies. We
therefore extrapolated the extinction curve to the X-ray region
using the average of the Milky Way R = 3.1 and R = 5.5 curves
calculated by B. Draine. These is based on the carbonaceous -
silicate dust model of Weingartner \& Draine (2001) and Li \& Draine
(2001). The graphite and silicate optical properties are given in
Draine (2003a,b), and include structure around X-ray absorption
edges.

\section{Adopted Spectrum}

The only remaining uncertainty in our derived spectrum (see Fig.\@
6) is the ``EUV gap''.  After our reddening corrections the
$\nu$F$\nu$ spectrum is rising on either side of the gap.  We
interpolated across the gap by fitting a 6th-order polynomial to the
points on either side of the gap.  This gives a good fit on either
side of the gap and we consider this to be the most likely
interpolation through the gap. We also did a simple linear
interpolation across the gap and this is also shown in Fig.\@ 6.
The latter is most unlikely to be the correct interpolation, but
this extreme case shows the maximum uncertainties introduced by our
extrapolation. For convenience we also tabulate our final continuum
in Table 2. This gives the continuum as it would be seen from the
earth with no extinction and no host galaxy light contamination.


\begin{figure}
\plotone{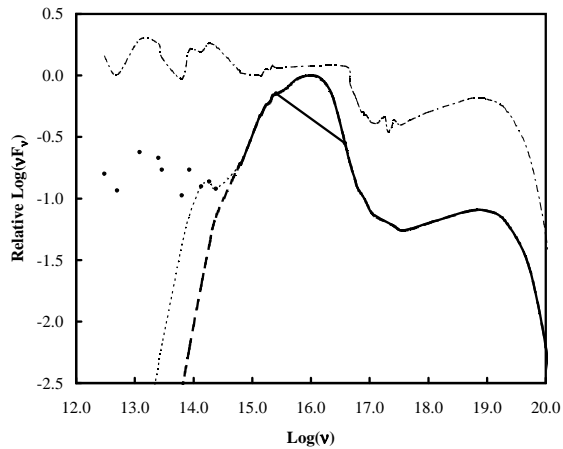} \caption{Estimated spectral energy distribution of
NGC~5548 around JD 2449120.  The upper dashed line represents the
raw observed or estimated continuum before any reddening or host
galaxy corrections. The thick solid line and the dots show the
corrected observed continuum from the optical to the X-ray region
including the interpolations in the EUV gap.  This represents the
SED as it would be seen from the earth with no extinction and no
host galaxy light. The straight line shows the gap. The IR
observations are shown as dots, while the lower dashed lines show
the IR at 7 ld (thick dashed) and at 80 ld (thin dashed). The
vertical scale is arbitrary.}
\end{figure}

\begin{center}
\begin{deluxetable}{crcr}
\tablewidth{0pt} \tablecaption{De-reddened Continuum Fluxes}
\tablehead{ Log($\nu$)& \colhead{Log($\nu
F_{\nu}$)\tablenotemark{a}} & Log($\nu$)& \colhead{Log($\nu
F_{\nu}$)\tablenotemark{a}}} \startdata
12.5 & -10.19 & 16.5 & -9.79 \\
13.0 & -10.07 & 17.0 & -10.47 \\
13.5 & -10.17 & 17.5 & -10.63 \\
14.0 & -10.20 & 18.0 & -10.81 \\
14.5 & -10.25 & 18.5 & -10.51 \\
15.0 & -9.87 & 19.0 & -10.52 \\
15.5 & -9.51 & 19.5 & -10.74 \\
16.0 & -9.38 & 20.0 & -11.79 \\
\enddata
\tablenotetext{a}{$\nu$F$\nu$ in ergs $s^{-1}$  $cm^{-2}$ at the
earth}
\end{deluxetable}
\end{center}

\section{Emission-Line Fluxes}


 In Table 3 we give the observed emission-line fluxes for
the highest state during the period we consider.  The Ly~$\alpha$, C
IV, and CIII] fluxes are from Korista et al.\@ (1995), the H$\beta$
flux is from Wanders \& Peterson (1996)\footnote{See Footnote 1},
and the He II $\lambda$1640 flux is from Bottorff et al.\@ (2002).
As Bottorff et al.\@ discuss, there is a large uncertainty ($\sim
30$\%) in this flux. There are also even larger uncertainties in
measuring He II $\lambda$4686 (see Appendix in Bottorff et al.\@
2002) and, in addition, the He II $\lambda$4686 line was not
measured in the 1993 campaign.  The value given in Table 1 was
scaled from the 1989 campaign results. H$\alpha$ was also not
observed in the 1993 campaign so the strength given in Table 1 comes
from assuming that H$\alpha$/H$\beta$ $\approx 4.1$ (Shapavalova et
al.\@ 2004). The slight variations in this ratio with time (see
Shapavalov et al.\@) are unimportant for our analysis.  He I
$\lambda$ 5876 was also not observed during the 1993 campaign, so in
Table 3 we have simply scaled it from H$\beta$ using the median He I
$\lambda$5876/H$\beta$ ratio from Osterbrock (1977).  As can be seen
from Shapavalova et al.\@ (2004) He I $\lambda$5876 varies
substantially. Table 3 also gives the fluxes corrected for Galactic
and internal reddening as discussed above.

\begin{deluxetable}{lcrrr}
\tablewidth{0pt} \tablecaption{Emission-Line Fluxes} \tablehead{
\colhead{Line} & \colhead{} & \colhead{Obsd.} & \colhead{MW de-red}
& \colhead{Int. de-red}} \startdata
Ly$\alpha$ & $\lambda$1216& 480& 600 & 2300\\
N V & $\lambda$1240 & 93& 120 & 450\\
O I & $\lambda$1304 & 30 & 40 & 140\\
O IV] + Si IV & $\lambda$1400 & 90 & 110 & 430\\
C IV & $\lambda$1549 & 730 & 870  & 3200 \\
He II & $\lambda$1640 & 75 & 89 & 320\\
C III] & $\lambda$1909 & 135 & 160 & 550\\
Mg II & $\lambda$2798 & 130 & 150 & 400\\
He II & $\lambda$4686 & 15 & 16 & 30\\
H$\beta$ & $\lambda$4861 & 62 & 70 & 120\\
He I & $\lambda$5876 & 11 & 12 & 18 \\
H$\alpha$ & $\lambda$6563 & 250 & 270 & 390\\
\enddata
\end{deluxetable}


\section{Reddening from lines}

Emission-line ratios can potentially give an independent check on
reddening derived from the continuum.  The He~II recombination lines
give the most intrinsically reliable line ratios, since they are
mostly independent of physical conditions, but, as has been
extensively discussed by Bottorff et al.\@ (2002), measuring the He
II $\lambda$1640/$\lambda$6886 ratio is very difficult (see their
Appendix), so while the He~II fluxes in Table 3 are consistent with
the adopted reddening, the He II line ratio does not provide a
reliable constraint.

The Balmer decrement is sensitive to physical conditions and varies
both with projected velocity (see Snedden \& Gaskell 2004, 2007a,b)
and with time, but the relative integrated H$\alpha$/H$\beta$ ratio
is an approximate indicator of reddening (e.g., De Zotti \& Gaskell
1985). Shapavalova et al.\@ (2004) find H$\alpha$/H$\beta$ varying
from 3.5 to 4.3, which is not too different from Osterbrock (1977).
If we adopt an unreddened decrement of 3.0, then the NGC~5548
decrements give E(B-V) $\sim 0.17$ -- 0.36, both which are
consistent with the reddening we get from the continuum.

\section{Photoionization Modelling}

After determination of the continuum we ran models using the
photoionization code CLOUDY, version 96.01 (Ferland et al.\@ 1998)
to determine the emission-line spectrum. In order to match the
source-to-cloud distances found from reverberation mapping (Korista
et al.\@ 1995) and our estimated 80 ld inner edge of the dusty
torus, we ran CLOUDY models corresponding to source distances of 7,
14, 21, 30, 50, and 80 ld. The continua at each of these radii were
calculated as described above.  Since there will certainly be a
range of densities in the BLR, we ran models with hydrogen
densities, $n_H$, of $10^{9.5}$, $10^{10.5}$, and $10^{11.5}$
cm$^{-3}$.  Because they are located at the very centers of
galaxies, the metalicities of AGNs are almost always supersolar
(Hamann \& Ferland 1999), so we adopted a metal abundance of three
times solar. All models run were completely optically thick models
and the calculations were continued far deeper into the clouds than
any of the lines we are interested in were being produced.

In addition to these models we ran additional models to investigate
the effect of not reducing the IR emission, and of varying the metal
abundances from solar to six times solar. We also investigated the
effect of significant (1000 km s$^{-1}$) (non-dissipative) internal
turbulence (see Bottorff et al.\@ 2000 and Bottorff \& Ferland
2002).


\begin{deluxetable*}{lrrrrrrrrrrrr}
\tablewidth{0pt} \tablecaption{Covering Factors} \tablehead{
\colhead{Model} & \colhead{log($n_H$)} & \colhead{$r$(ld)} &
\colhead{Ly$\alpha$} & \colhead{N V} & \colhead{O I } &
\colhead{$\lambda$1400}  & \colhead{C IV} & \colhead{He II} &
\colhead{C III]} & \colhead{Mg II}& \colhead{H$\beta$} & \colhead{He
I}}
\startdata
Adopted SED & 9.5 & 7 & 74\% & 157\% & & 192\% & 223\% & 25\% & 77\% & 212\% & 426\%\\
 & 9.5 & 14 & 28\% & 86\% & 417\% & 113\% & 131\% & 17\% & 32\% & 306\%  & 80\% & 123\% \\
 & 9.5 & 21 & 17\% & 60\% & 196\% & 85\% & 106\% & 19\% & 21\% & 119\%  & 44\% & 66\% \\
 & 9.5 & 30 & 13\% & 43\% & 135\% & 70\% & 83\% & 21\% & 16\% & 68\% & 28\% & 43\% \\
 & 9.5 & 50 & 10\% & 28\% & 106\% & 54\% & 55\% & 22\% & 11\% & 28\% & 18\% & 29\% \\
 & 9.5 & 80 & 9\% & 21\% & 87\% & 44\% & 35\% & 24\% & 10\% & 15\% & 15\% & 30\% \\
 &  &  &  &  &  &  &  &  &  &  &  &  \\
 & 10.5 & 7 & 35\% & 43\% & 173\% & 77\% & 145\% & 15\% & 81\% & 526\%  & 218\% & 181\% \\
 & 10.5 & 14 & 19\% & 28\% & 36\% & 79\% & 101\% & 21\% & 60\% & 200\%  & 97\% & 75\% \\
 & 10.5 & 21 & 15\% & 22\% & 31\% & 64\% & 70\% & 22\% & 50\% & 101\%  & 57\% & 50\% \\
 & 10.5 & 30 & 13\% & 19\% & 34\% & 51\% & 48\% & 22\% & 46\% & 56\% & 39\% & 35\% \\
 & 10.5 & 50 & 11\% & 24\% & 47\% & 37\% & 29\% & 24\% & 45\% & 23\% & 24\% & 22\% \\
 & 10.5 & 80 & 10\% & 59\% & 48\% & 31\% & 24\% & 26\% & 44\% & 9\% & 17\% & 14\% \\
 &  &  &  &  &  &  &  &  &  &  &  &  \\
 & 11.5 & 7 & 37\% & 20\% & 66\% & 91\% & 120\% & 26\% & 320\% & 290\%  & 303\% & 159\% \\
 & 11.5 & 14 & 27\% & 19\% & 50\% & 50\% & 60\% & 23\% & 364\% & 100\%  & 128\% & 71\%  \\
 & 11.5 & 21 & 22\% & 27\% & 47\% & 35\% & 43\% & 22\% & 323\% & 50\% & 77\% & 42\% \\
 & 11.5 & 30 & 20\% & 53\% & 49\% & 29\% & 39\% & 23\% & 278\% & 28\% & 50\% & 27\% \\
 & 11.5 & 50 & 17\% & 256\% & 49\% & 29\% & 44\% & 24\% & 244\% & 12\% & 29\% & 15\% \\
 & 11.5 & 80 & 15\% &       & 39\% & 49\% & 75\% & 26\% & 280\% & 6\%  & 17\% & 10\% \\
 &  &  &  &  &  &  &  &  &  &  &  &    \\
SED &  &  &  &  &  &  &  &  &  &  &  &    \\
Observed & 10.5 & 21 & 64\% & 55\% & 147\% & 69\% & 71\% & 100\% & 77\% & 51\% & 115\% & 107\%\\
Adopted & 10.5 & 21 & 15\% & 22\% & 31\% & 64\% & 70\% & 22\% & 50\% & 101\%  & 57\% & 50\% \\
Linear & 10.5 & 21 & 22\% & 31\% & 38\% & 84\% & 74\% & 41\% & 75\% & 100\% & 65\% & 62\% \\
Adopted  & 10.5 & 7 & 35\% & 43\% & 173\% & 77\% & 145\% & 15\% & 81\% & 526\% & 218\% & 181\% \\
Extra IR & 10.5 & 7 & 36\% & 32\% & 172\% & 55\% & 103\% & 17\% & 67\% & 437\% & 294\% & 206\% \\
 &  &  &  &  &  &  &  &  &  &  &  &   \\
Fe/H &  &  &  &  &  &  &  &  &  &  &  &   \\
Solar & 10.5 & 7 & 36\% & 80\% & 723\% & 117\% & 139\% & 12\% & 104\% & 908\% & 207\% & 194\% \\
3 solar & 10.5 & 7 & 35\% & 43\% & 173\% & 77\% & 145\% & 15\% & 81\% & 526\% & 218\% & 181\%  \\
6 solar & 10.5 & 7 & 36\% & 39\% & 92\% & 53\% & 135\% & 20\% & 68\% & 332\%  & 225\% & 174\%  \\
3 solar & 10.5 & 14 & 19\% & 28\% & 36\% & 79\% & 101\% & 21\% & 60\% & 200\%  & 97\% & 75\% \\
6 solar & 10.5 & 14 & 21\% & 28\% & 23\% & 58\% & 108\% & 27\% & 49\% & 120\%  & 97\% & 79\% \\
 &  &  &  &  &  &  &  &  &  &  &  &   \\
Turbulence &  &  &  &  &  &  &  &  &  &  &  &  \\
No turb. & 10.5 & 7 & 35\% & 43\% & 173\% & 77\% & 145\% & 15\% & 81\% & 526\%  & 218\% & 181\% \\
Turb. & 10.5 & 7 & 10\% & 25\% & 107\% & 37\% & 44\% & 17\% & 16\% & 33\% & 12\% & 21\% \\
No Turb & 10.5 & 21 & 15\% & 22\% & 31\% & 64\% & 70\% & 22\% & 50\% & 101\%  & 57\% & 50\% \\
Turb. & 10.5 & 21 & 8\% & 16\% & 58\% & 20\% & 27\% & 21\% & 26\% & 12\% & 11\% & 25\% \\
 &  &  &  &  &  &  &  &  &  &  &  &  \\
LOC Model &  &  & 17\% & 34\% & 61\% & 51\% & 56\% & 22\% & 37\% & 28\% & 36\% & 13\% \\

\enddata
\end{deluxetable*}

\section{Broad-Line Covering Factors}

In Table 4 we present the covering factors needed to produce the
observed line fluxes.  In order to show the effect of the radial
distribution of the gas for each line we give the covering factors
needed as a function of radius under the assumption that each line
is only produced in gas at this radius.  We show the effects of
varying density, source distance, metal abundance, and continuum
shape, and of adding turbulence. As well as giving covering factors
using our standard continuum (``adopted'') we also show the effects
of using the simple linear interpolation across the EUV gap
(``linear''), of using just the observed SED (``observed'') with no
corrections at all, and of not allowing for the BLR being far from
the IR emission (``Extra IR''). As is to be expected, changing the
EUV continuum shape has the biggest effect on the He II
recombination lines (see, for example, Mathews \& Ferland 1987).
Adding extra IR heating increases the strengths of all
collisionally-excited lines.

Looking at the density and radius dependencies of the covering
factors for our standard model it can be seen that, as is well known
(see, for example, Laor 2007), the production efficiency of most
lines decreases at high ionizing flux levels. The efficiency
generally increases with density for many lines except where
lowering the resultant lowering of the ionization reduces the line
flux (e.g., N V), or a line is collisionally suppressed (e.g., C
III]).

The observed integrated line fluxes of an AGN are the sums of the
contributions from all clouds at all densities and all radii (cf.
the ``LOC'' model of Baldwin et al.\@ 1995 and Korista \& Goad
2000). In the bottom line of Table 4 we give the covering factors
for an integrated model based on our standard continuum model, with
a uniform range of densities from $10^{9.5}$ to $10^{11.5}$
cm$^{-3}$, and with uniform covering factor as a function of radius.
There is generally good agreement in the resulting covering factors
which range from 17--56\% (ignoring He I $\lambda$5876 for which the
observed flux is highly uncertain) and with an average is 40\%. This
average covering factor can be lowered somewhat by weighting the
density distribution towards higher densities, and by having the
covering factor increase with radius, but it can be seen from Table
4 that an average integrated covering factor of $< 20$\% is not
attainable. The only way of lowering the average covering factor to
$< 20$\% is by including substantial internal turbulence.  Further
lowering of the covering factor can be achieved by making the
turbulence dissipative to provide additional heating for the
low-ionization lines (Bottorff \& Ferland 2002).

\section{The Torus Covering Factor}

The dust covering factor, $\Omega_{dust}$, can be found from the
ratio of the bolometric luminosity, $L_{bol}$, to the luminosity, in
the mid-infrared, $L_{IR}$.  Since we have the complete spectrum for
NGC~5548 we can perform a direct integration to obtain $L_{Bol}$. We
show the results for different continuum assumptions in Table 5. The
observed case ignores reddening and the host galaxy correction.  The
``low big bump'' case is the simple linear interpolation across the
EUV gap and the ``high big bump'' case is the more likely
interpolation (see Fig.\@ 6). The covering factors given in Table 5
assume that the torus has an unobstructed view of the higher-energy
continuum.  We will argue below that this is probably not the case.

\begin{center}
\begin{deluxetable}{lc}
\tablewidth{0pt} \tablecaption{ESTIMATED DUST COVERING FACTORS}
\tablehead{\colhead{Continuum Shape} & \colhead{$\Omega_{dust}$}}
\startdata
Observed (uncorrected) & 61\% \\
Internally de-reddened with low big bump & 25\% \\
Internally de-reddened with high big bump & 19\% \\
\enddata
\end{deluxetable}
\end{center}

\section{Variability Transfer Functions}

The radial distribution of material in the BLR can only be inferred
indirectly from reverberation mapping (see Peterson 2006 for a
recent review). The most widely used measure of the size of the BLRs
of AGNs is the ``lag'' given by the centroid of the
cross-correlation function (Gaskell \& Sparke 1986, Gaskell \&
Peterson 1987), but this has always been known to be biased towards
gas at small radii. A better indication of the distribution of the
gas is given by the transfer function $\Psi(\tau)$, which is the
response of the system at a delay $\tau$ to a delta function at
$\tau = 0$. A detailed comparison of observed and theoretical
transfer functions is beyond the scope of this paper and will be
given elsewhere, but in Figs.\@ 7 -- 9 we compare observed transfer
functions of C IV and H$\beta$ with simple theoretical transfer
functions based on the parameters of our photoionization modelling.

Recovering the $\Psi(\tau)$ from observed continuum and line-flux
time series is difficult.  In addition to uncertainties arising from
noise in the data (which are most significant for weak lines like
those of He~II), the derived $\Psi(\tau)$ depends on the sampling,
the method used, and assumptions that always need to be made to
recover $\Psi(\tau)$ in practice. Assumptions need to be made about
the smoothness of $\Psi(\tau)$ and boundary conditions including the
contribution of relatively non-varying components. It also needs to
be recognized that some methods (e.g., the maximum-entropy method)
only allow $\Psi(\tau)$ to be positive, so noise in $\Psi(\tau)$
gives spurious positive contributions at large delays. Since AGN
variations are commonly quasi-cyclical on the timescale of almost
all monitoring campaigns, aliasing at multiples of the quasi-period
of the continuum variations will also make spurious contributions to
$\Psi(\tau)$ at large delays. Because of these methodolical
differences, a variety of different transfer functions have been
published from the same data sets even for strong lines.

As is well known, $\Psi(\tau)$ for a thin spherical shell of radius
$r$ is a boxcar from $\tau = 0$ to $\tau = 2 r/c$. The transfer
functions for other spherically-symmetric distributions can be
constructed by adding such boxcar functions.  For example, a filled
sphere has $\Psi(\tau)$ declining monotonically from a peak at $\tau
= 0$. A hollow sphere of inner radius $r_{in}$ has $\Psi(\tau)$
constant until $\tau = 2 r_{in}/c$.  If the clouds emit
isotropically, the only way to avoid $\Psi(\tau)$ having a maximum
at $\tau = 0$ is to have a non-spherical geometry. Most type-1 AGNs
are observed close to pole on so we modelled non-spherically
symmetric distributions as spheres with polar cones cut out of them
viewed from on axis.  A single thin sphere with a cap cut out around
each of the poles gives $\Psi(\tau)$ as a box car of width $\pm$
sin($\theta$)$r/c$ centered on $r/c$, where $\theta$ is the maximum
angle the clouds subtended above the equatorial plane.  $\Psi(\tau)$
for radially-thick distributions is simply the sum of these box
cars. We calculated simple theoretical AGN line transfer functions,
smoothed to the observed resolution, for clouds in a distribution
which is truncated by cones on the symmetry axis, but is otherwise
spherically symmetric. $\Psi(\tau)$ for geometries where the
observer is slightly off axis, as will be the case for most AGNs,
are mostly indistinguishable from on-axis cases. Off-axis cases can
be divided into two classes: when the observer is inside the cone,
and when he or she is outside the cone.  If the opening of the cone
is wide, the former case will be indistinguishable from the on-axis
case, and if the opening of the cone is narrow, the latter case will
be indistinguishable from spherical symmetry.

Our models have only three main parameters: the half-opening angle
of the polar cone, $90\deg - \theta$, the outer radius, and the
radial variation of responsivity which we parameterized as a simple
power laws. The radial variation of the responsivity depends on
three things: the response of a particular line to changes in the
ionizing continuum, the emissivity per cloud as a function of
radius, and the dependence of the covering factor (number of clouds)
as a function of radius. The functional form of the response of the
flux in a given line from a cloud to changes in the continuum is
expected to remain nearly constant over the range of conditions
considered. The radial variation of responsivity thus depends on the
change in emissivity, which is given by our photoionization models,
and the unknown dependence of the covering factor on radius.  For
simplicity we assumed that the covering factor, which has to be the
same for all lines coming from a given type of cloud, has no radial
gradient, and that the radial variation in responsivity thus depends
solely on the radial variation of emissivity given by the
photoionization models. The remaining parameters for each transfer
function are the opening angle and the outer radius. Although these
assumptions are very simple, they nonetheless give useful insights
into the behavior of transfer functions.

\begin{figure}
\plotone{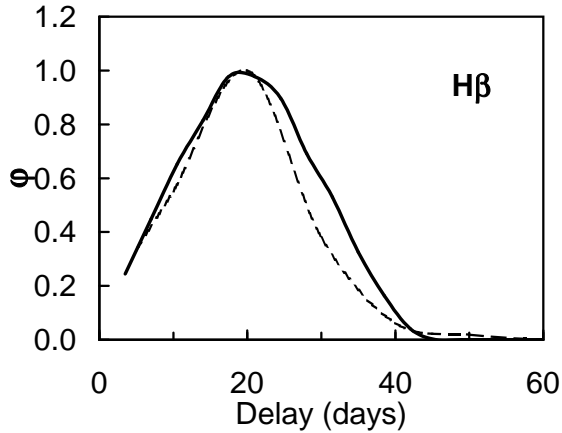} \caption{Theoretical (solid) and observed (dashed)
transfer functions for H$\beta$ in NGC 5548.  The observed transfer
function is from Horne, Welsh, \& Peterson (1991).}
\end{figure}

\subsection{The H$\beta$ Transfer Function}

The photoionization models show that the emissivity of H$\beta$
rises as $\sim r^{+1}$.  Horne, Welsh, \ Peterson (1991) found that
$\Psi(\tau)_{H\beta}$ (which we show as a dotted line in Fig.\@ 7)
does not peak at zero delay.  Cacklett \& Horne (2006) show that
over the course of the 13-year IAW monitoring of NGC~5548
$\Psi(\tau)_{H\beta}$ maintained the same shape regardless of the
mean flux level of NGC~5548. As outlined above, such a transfer
function demands a predominantly non-spherical
distribution\footnote{Ferland et al.\@ 1992 show how a spherical
distribution can produce the H$\beta$ transfer function if the
emission is preferentially back towards the source, but this is
unlikely since the Ly $\alpha$ transfer function in NGC~5548 is
incompatible with such asymmetric emission.  Also, the double-humped
H$\beta$ profile is independent evidence against spherical
symmetry.} seen from within the polar opening. As can be seen in
Fig.\@ 7, we can readily reproduce the shape of
$\Psi(\tau)_{H\beta}$ with our simple models. The symmetry of
$\Psi(\tau)_{H\beta}$ requires $\theta \lesssim 25\deg$. This
corresponds to a covering factor of 40\% or less, and thus is in
good agreement with the results in Table 4. For such a distribution,
the position of the peak of $\Psi(\tau)$ depends only on the outer
radius of the distribution. The outer edge of the BLR, $r_{out}$ is
expected to be the dust sublimation radius (Netzer \& Laor 1993,
Laor 2007) which is quite well defined. From consideration of IR
reverberation mapping in section 2.2 we estimated  that $r_{dust}
\thickapprox 80$ ld, but we can get a more precise estimate of
$r_{out}$ by fitting $\Psi(\tau)_{H\beta}$.  At the optical flux
level of NGC~5548 during the high state we are considering, Cackett
\& Horne (2006) find the peak of the $\Psi(\tau)_{H\beta}$ to be at
19 ld, with lower and upper quartile uncertainties of -2 and +5 ld
respectively. This is consistent with the lag determined by the
Korista et al.\@ (1995). Fitting our model transfer function gives
$r_{out} = 60$ ld with the same fractional uncertainty as in the
position of the observed peak of $\Psi(\tau)_{H\beta}$.

\subsection{The He II Transfer Function}

The transfer function of He II was determined by Krolik et al.\@
(1991).  It is not as well determined as that of H$\beta$, but, as
can be seen in Fig.\@ 8, it shows a strong peak at zero delay. The
emissivity of He II, unlike that of H$\beta$, declines slowly with
radius as $\sim r^{-0.4}$.  In our models this declining emissivity
automatically gives $\Psi(\tau)$ peaking towards zero delay.  The
radical difference in the shape of $\Psi(\tau)_{H\beta}$ and
$\Psi(\tau)_{HeII}$ thus primarily arises from photoionization
physics (how clouds emit as a function of incident flux) rather than
a profoundly different spatial distribution. Nonetheless, there must
be some difference in the geometries of the clouds dominating the
H$\beta$ emission and those dominating the He~II emission.   If we
are to keep the same flattened distribution derived from the
$\Psi(\tau)_{H\beta}$ then $\Psi(\tau)_{HeII}$ declines too slowly
with increasing delay.  To make our model match the observed
$\Psi(\tau)_{HeII}$ requires us to reduce $r_{out}$ to $\sim 30$
days.  However, if the clouds are directly exposed to the ionizing
continuum (as in our CLOUDY models), we still have He~II emission
wherever there is H$\beta$ emission, and the He~II emitting cloud
distribution would not be truncated more than the H$\beta$ emitting
cloud distribution. In this case another way of matching the shape
of the He~II transfer function (see Fig.\@ 8) is to increase the
covering factor. For pole-on viewing a covering factor of over 90\%
is needed ($\theta > 60\deg$). A more likely possibility is that the
covering factor is only slightly greater than for H$\beta$ (and
hence in agreement with the results of Table 4), but we are now
outside the hollow polar cone.

\begin{figure}
\plotone{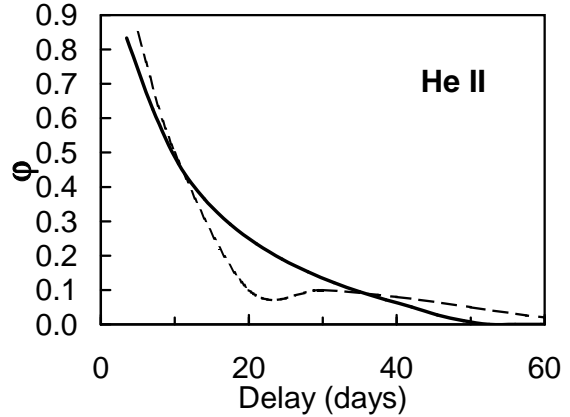} \caption{Theoretical (solid) and observed (dashed)
transfer functions for He~II $\lambda$1640 in NGC 5548. The observed
transfer function is from Krolik et al.\@ (1991).}
\end{figure}

\subsection{The C IV Covering Factor}

In Fig.\@ 9 we show  $\Psi(\tau)_{CIV}$ estimated at two epochs by
two different methods by Wanders et al.\@ (1995) and Horne et al.\@
(1991). Additional estimated C~IV transfer functions can be found in
these papers and also in Done \& Krolik (1996). There is general
agreement in $\Psi(\tau)_{CIV}$ for $\tau < 20$ d for both epochs,
but more uncertainty for $\tau > 30$ d. For C~IV the radial
dependence of the emissivity is intermediate between that of
H$\beta$ and He~II.  In the model shown in Fig.\@ 9 we adopt a flat
radial dependence.  If we again assume $r_{out} \sim 60$d, as
estimated from $\Psi(\tau)_{H\beta}$, then to fit the observed
$\Psi(\tau)_{CIV}$ we need a high covering factor as for He~II.
Again this could really be a lower covering factor with us looking
through the equatorial clouds.  This is quite likely to be the case
for NGC~5548 since the appearance of a double-humped, disk-like
H$\beta$ profile implies that we are not viewing it pole-on. It is
also well known (see Arav et al.\@ 2003, Steenbrugge et al.\@ 2003
Steenbrugge et al.\@ 2005, and refs. therein) that there is C~IV
absorption in NGC~5548.

\begin{figure}
\plotone{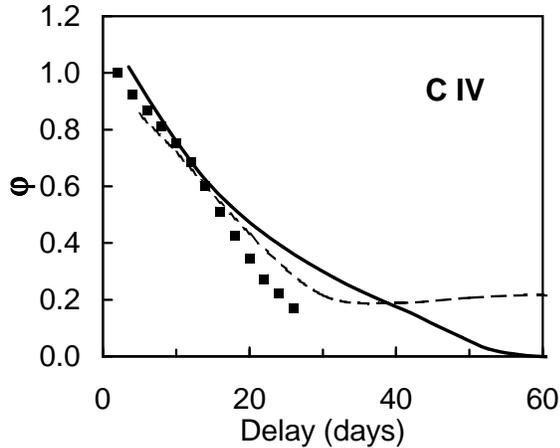} \caption{Theoretical and observed transfer
functions for H$\beta$ in NGC 5548.  The theoretical transfer
function (see text) is shown as a solid line.  The squares are the
transfer function derived with the SOLA method from the 1993
campaign (Wanders et al.\@ 1995) and the dashed line is the transfer
function derived from the 1989-90 campaign by Horne et al.\@ (1991)
using the maximum entropy method.}
\end{figure}

\section{DISCUSSION}

\subsection{Is There an Energy-Budget Problem?}

We have considered in detail the overall SED of NGC~5548 during the
high state at JD 2449120 and argue that although not all spectral
regions were observed, we can be fairly confident of the overall
SED.  The main assumption still to be confirmed is the suggestion of
Gaskell \& Benker (2007) that the apparent turndown in the SED below
Ly $\alpha$ is not intrinsic but results from a small amount of
SMC-like dust.  If this is incorrect, it only has a slight effect on
our estimated covering factors.

The He~II recombination lines, which are a very good indicator of
covering factor since they are very insensitive to physical
conditions, only require a covering factor of $\sim 25$\% over a
wide range of conditions with our adopted model. Other lines depend
on physical conditions more but in our integrated (``LOC'') model
only two lines, O~I $\lambda$1304 and C IV require covering factors
of more than 50\%. Only the He~II flux depends significantly on our
assumption that there is a small amount of SMC-like reddening.  O~I
$\lambda$1304 is very sensitive to the metalicity and a slight
increase in metalicity is sufficient to bring the O~I covering
factor in line with the other lines.  High metalicity also lowers
the covering factors needed for N~V, the O~IV] + Si~IV $\lambda$1400
blend, and, to a lesser extent, C III]. Increasing the metalicity
does not help reduce the C~IV covering factor since it is a major
coolant.  What does help reduce the C~IV covering factor most is an
increased contribution of higher density gas. Higher density gas
generally increases the emission efficiency of most lines.

From Table 4 it can be seen that with what we consider to be the
best continuum, most of the lines can be explained by a covering
factor of $\sim 40$\% or somewhat less.  This is not significantly
lower than covering factors obtained in the past with weaker EUV
continua (see, for example, Shields \& Ferland 1993; Korista et
al.\@ 1998; Goad \& Koratkar 1998; Kaspi \& Netzer 1999; Korista \&
Goad 2000), so our main point is that even though we have argued
that because of reddening there are more photons in the EUV
continuum than hitherto recognized, {\it is it not possible to lower
the covering factor significantly.}  In particular, it is not
possible to lower it to the level required by the Lyman limit
absorption constraints.  The reason why reddening does not solve the
problem is that the lines have to be de-reddened as well as the
continuum.

Independent support for the BLR covering factor we derive comes from
the Fe K$\alpha$ line equivalent width.  Chiang et al.\@ (2000)
derive a Fe K$\alpha$ covering factor of 20--25\% for NGC~5548.

While all lines can be explained with covering factors significantly
less than 100\%, and so there is formally no ``energy-budget''
problem, we are unable to lower the necessary covering factors by an
order of magnitude. In the old scenario of clouds uniformly covering
the central source, a covering factor of $\sim 40$\% (or even 20\%)
would still be a problem because Lyman continuum absorption is
certainly not detected in 40\% of AGNs.  We believe that the
solution to this problem, as proposed by Maiolino et al.\@ (2001c),
is that the clouds do {\it not} cover the source uniformly. Instead,
they are in an aximuthally symmetric distribution and we are looking
in through a hole at the pole. Important support for this comes from
the line-continuum variability transfer functions discussed above.
We have shown that the geometry implied by the well-measured
H$\beta$ transfer function is in good excellent quantitative
agreement with this scenario. For other lines (e.g., C~IV and He~II)
a less flattened distribution is needed. Additional support for a
flattened geometry comes from the alignment of the polarization
vector parallel to the axis of symmetry (see discussion in Goosmann
\& Gaskell 2007), and from polarization reverberation (Gaskell et
al.\@ 2007).

Since our deduced covering factors can be reconciled with line
transfer functions for near pole-on viewing angles, we do not
believe that there is an energy-budget problem for NGC~5548.  This
also makes some previously suggested solutions to the problem
unnecessary. Given a flattened geometry there is no {\it need} for
additional heating (Dumont et al.\@ 1998) or for a large amount of
turbulence (Bottorff et al.\@ 2000, Bottorff \& Ferland 2002) to
reduce the covering factor, although such factors could still be
relevant.

\subsection{A New BLR Model}

If the majority of the BLR gas is located within $\pm 25\deg$ of the
equatorial plane, and the overall covering factor $\Omega/4\pi
\thickapprox 0.4$, then {\it photons leaving the central source in a
direction close to the equatorial plane have a close to 100\%
probability of striking BLR gas.}  By the time one gets to the inner
edge of the dusty torus the spectrum has been substantially modified
by passing through the inner BLR.  Such modification of the spectrum
has already been considered by Holt et al.\@ (1980) and modelled for
NGC~4151 by Ferland \& Mushotzky (1982). The effect of shielding is
similar to taking a standard cloud, cutting it in half in a plane
perpendicular to the direction of the incoming photons and moving
the outer part of the cloud further away.  The BLR scenario we
require resembles the geometry considered by MacAlpine (1972).  In
Fig.\@ 10 we contrast our proposed shielded BLR geometry with the
widely-used geometry where shielding is assumed to be negligible.

\begin{figure}
\plotone{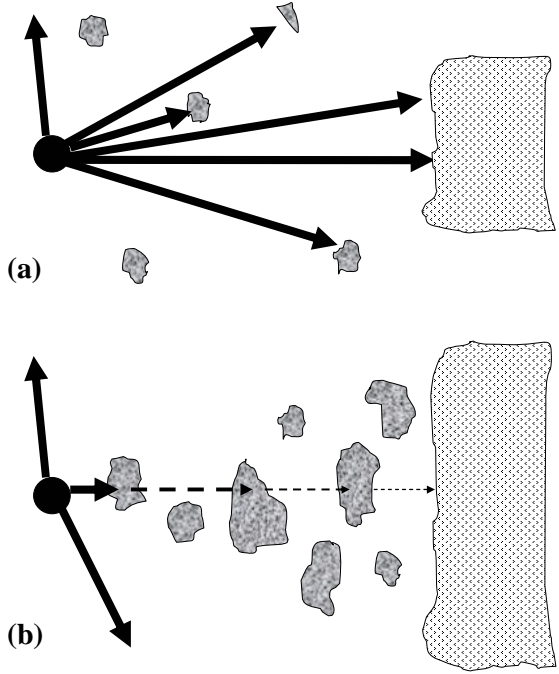} \caption{Schematic representation of BLR
geometries.  (a) Traditional model where BLR clouds are distributed
quasi-spherically and mostly do not shield distant clouds or the
torus. (b) Shielding model where the clouds are concentrated in the
equatorial plane. The torus and distant clouds see the central
source shielded by clouds closer to the center. (In both sketches
the BLR clouds are not shown to scale and the number of clouds has
been reduced by many orders of magnitude.) }
\end{figure}

As discussed in 12.2 and 12.3, the differences in the transfer
functions of C~IV and H$\beta$ require either that the
high-ionization emission comes from a smaller region, or that it has
a higher covering factor, or both.  It has in fact long been argued
that there are effectively two separate types of cloud:
high-ionization clouds and low-ionization clouds (Gaskell 1987;
Collin-Souffrin \& Lasota 1988; Gaskell 2000). Not only are
differences in radii implied by reverberation mapping, but the
higher velocity dispersions of high-ionization lines (Osterbrock \&
Shuder 1982; Shuder 1982) show that the high-ionization gas is
substantially closer (see Peterson \& Wandel 1999).  The
blueshifting of high-ionization lines (Gaskell 1982) requires that
the gas producing the high-ionization lines is spatially separated
from the low-ionization gas.  Furthermore, for NGC~5548 and other
objects we find that the velocity dependence of physical conditions
is different for the high- and low-ionization gas (Snedden \&
Gaskell 2004; Nazarova et al al. 2004, 2006; Snedden \& Gaskell
2007a,b).

There have been attempts (notably by Korista \& Goad 2000) to model
this stratification in detail with a continuous distribution of
cloud properties as in the LOC model, but there is a major
difficulty in quantitatively matching the range of observed lags and
line widths. As Korista \& Goad (2000) point out, there are problems
at both the high- and low-ionization ends: He~II and N~V have much
$smaller$ lags than predicted, while C~III] and Mg~II have
significantly $longer$ lags than predicted.  He~II presents another
problem.  The LOC models (Bottorff et al.\@ 2002) predict that the
radius of the He~II emitting region should be only 10\% smaller than
the C~IV emitting region, but as Bottorff et al.\@ point out, the
observations for NGC~5548 (in both 1989 and 1993), and for NGC 3783
(Reichert et al.\@ 1994), Fairall 9 (Rodriguez-Pascual et al.\@
1997), NGC~7469 (Wanders et al.\@ 1997), and 3C~390.3 (O'Brien et
al.\@ 1998) $all$ give a lag of He~II $\lambda$1640 three or four
times smaller than the C~IV lag.\footnote{Bottorff et al.\@ (2002)
also point out a problem with He~II $\lambda$4686 appearing to give
a lag longer than that of He~II $\lambda$1640.  The explanation of
the longer He~II $\lambda$4686 delay is probably that a small amount
of contamination by Fe~II and the blue wing of H$\beta$ makes the
He~II $\lambda$4686 lag appear longer.  See Gaskell (2007).}  In
Table 6 we compare the observed lags for NGC~5548 (normalized to a
C~IV lag of 8 days) with the predictions of the LOC model
predictions.  The LOC predicted lags have also been scaled to a C~IV
lag of 8 days. It can be seen the LOC model only predicts a factor
of four variation in lag while the observations give three times
that.

\begin{center}
\begin{deluxetable}{lcccc}
\tablewidth{0pt} \tablecaption{MEASURED AND PREDICTED LAGS FOR NGC
5548} \tablehead{\colhead{Line} & &
\colhead{Observed\tablenotemark{a}} & \colhead{LOC\tablenotemark{b}}
& \colhead{Shielded}} \startdata
Ly $\alpha$         &   $\lambda$1216    &   12   &    8.5 &   12    \\
N V          &   $\lambda$1240    &    4\tablenotemark{c}   &    3.8 &    2.4 \\
O I          &   $\lambda$1304    &   40\tablenotemark{c}   &        &   43   \\
Si IV + O IV]&   $\lambda$1400    &   12   &    5.8 &   15  \\
C IV         &   $\lambda$1549    &    8   &    8.0 &    8   \\
He II        &   $\lambda$1640    &    3\tablenotemark{d}   &    7.3\tablenotemark{d} &    3.3 \\
C III]       &   $\lambda$1909    &   35   &   11.5 &   28  \\
Mg II        &   $\lambda$2798    &   $\geq 34$   &   16 &   28  \\
He II        &   $\lambda$4686    &    3\tablenotemark{e}   &    7.3\tablenotemark{d} &    3.4 \\
H$\beta$       &   $\lambda$4861    &   20\tablenotemark{f}   &        &   18    \\
He I         &   $\lambda$5876    &   13\tablenotemark{g}   &        &   16    \\
H$\alpha$      &   $\lambda$6563    &   22.5\tablenotemark{g} &        &   19    \\
\enddata
\tablenotetext{a}{From Clavel et al.\@ (1991) except as noted.}
\tablenotetext{b}{Based on Korista \& Goad (2000) except as noted.}
\tablenotetext{c}{From Krolik et al.\@ (1991).}
\tablenotetext{d}{From Bottorff et al.\@ (2002).}
\tablenotetext{e}{Set equal to $\lambda$1640 lag (see footnote 8).}
\tablenotetext{f}{From Peterson et al.\@ (1991).}
\tablenotetext{g}{Scaled from H$\beta$ using square of mean ratio of
FWHMs from Osterbrock \& Shuder (1982).}
\end{deluxetable}
\end{center}

The shielding model shown in Fig.\@ 10b naturally explains the
increased stratification. The relative order of ionization stages in
a photoionized cloud is very insensitive to changes in the density
or the ionizing flux; only the scale factor changes.  We can
therefore approximate a shielded distribution of clouds by taking a
single cloud in CLOUDY and pretending that it has been broken into
smaller pieces spread uniformly throughout the BLR.  We simply
linearly transform the depths into the cloud from the side facing
the ionizing source into radii away from the central source.  For
different lines of sight and for different densities, cloud sizes,
and filling factors the scale factor will be different, but the
order of stages of ionization remains the same.  The weighted mean
radii of maximum emission of each line will therefore remain in the
same ratios as the depths into a single cloud.  One important
difference from a single cloud is that photons can now escape from
the distribution much more easily.  We treated this in the same
manner as dissipationless turbulence.

To quantitatively test the shielded BLR model, we took the 21 ld,
$10^{10.5}$ cm$^{-3}$ CLOUDY model with a 1000 km s$^{-1}$
turbulence velocity, and scaled the radius of peak C~IV emission to
match the observed 8-day lag.  The only other free parameter is the
outer edge of the BLR which we took to be the 80 ld inner edge of
the torus. We show our predicted lags in Table 6.  As can be seen
this give a better quantitative agreement with the observed lags
($\pm 20$\%) than the LOC model.

\subsection{The Torus Covering Factor}

Netzer \& Laor (1993) proposed that the outer edge of the BLR is set
by the dust-sublimation radius.  Our modeling of transfer functions
requires a fixed outer radius of $\sim 60$d, and we have shown that
this is in agreement with the radius of the hot dust measured by IR
reverberation mapping.

We consider the most likely dust covering factor to be 19\% (see
section 11). This is lower than almost all the covering factors we
deduced for the various emission lines in Table 4 and would imply
that the dust has a flatter distribution than the gas.

The dust covering factor we find for NGC~5548 is similar to those of
other AGNs.  Maiolino et al.\@ (2007) discuss the dust covering
factors for a wide range of AGNs.  They estimate the dust covering
factor by assuming a ratio of $L_{Bol} = 7 \nu L_{\nu}(\lambda
5200)$. However, the study of Shang et al.\@ (2005) implies that
$L_{Bol} = 15 \nu L_{\nu}(\lambda 5200)$ is more appropriate, so
that the covering factor can be calculated from

\begin{equation}
\Omega \approx 0.18 \nu F_{\nu}(6.7\mu m) / F_{\nu}(5100 \AA)
\end{equation}

For NGC~5548 this gives $\Omega/4\pi \approx 0.18 $ which  is
comparable to dust covering factors Maiolino et al.\@ (2007) find
for AGNs of comparable luminosity if we correct for our updated
scaling of $\nu L_{\nu}(\lambda 5200)$ to $L_{Bol}$

Although the deduction of $\Omega_{dust}/4\pi \approx 0.2$ is
straight forward and consistent with the covering factors we deduced
for low-ionization lines, there is a problem: our NGC~5548 torus
covering factor and similarly estimated covering factors for other
AGNs are smaller than the torus covering factors implied by the
observed ratio of type-1/type-2 AGNs. The ratio of type-1/type-2
AGNs in hard X-ray samples (see Fig.\@ 7a of Maiolino et al.\@ 2007)
implies that at the luminosity of NGC~5548, $\Omega_{dust}/4\pi =
0.6$.  A further problem of having $\Omega_{torus} < \Omega_{BLR}$
is that there should be many more objects where we see strong
low-ionization absorption by the BLR (Maiolino et al.\@ 2001c).
While there are low-ionization broad absorption line AGNs, this
absorption is not arising from the main BLR because it is
blueshifted, and hence arising from outflowing gas. Because the
variability of the blue wings of broad emission lines does not lead
that of the red wings (Gaskell 1988; Koratkar \& Gaskell 1989,
Crenshaw \& Blackwell 1990) the main BLR is not outflowing but is
gravitationally bound. Observations indicate in fact that the BLR is
{\it inflowing} slightly (Gaskell, 1988, Koratkar \& Gaskell 1989,
Koratkar \& Gaskell 1991a,b, Korista et al.\@ 1995; Done \& Krolik
1995; Gaskell \& Snedden 1997, Welsh et al.\@ 2007).  The problem of
the lack of BLR absorption at the systemic velocity requires
$\Omega_{torus} < \Omega_{BLR}$

We believe that our proposed BLR geometry offers the solution to
these two problems: the BLR is almost certainly co-planar with the
torus, so the latter does {\it not} have an unobscured view of the
central source.  Therefore, the energy received by the torus is
substantially reduced, and hence the torus has to subtend a larger
solid angle to explain the observed IR luminosity.  If we assume
that all of the energy from the Lyman limit to 1 keV is absorbed,
the dust covering factor increases by a factor of 2.7 to 53\%, which
is consistent with the mean torus covering factor expected from the
ratio of type-1/type-2 AGNs at the luminosity of NGC~5548.

\subsection{The Inner Radius of the Torus}

BLR absorption can also help another problem.  Barvainis (1992) was
able to satisfactorily explain the $\sim 400$ d IR lag in Fairall 9
with dust sublimation temperatures of 1300 -- 2000 K. However,
Fairall 9 lies at the top of the IR lag-luminosity relationship
shown in Fig.\@ 1 and there is an order of magnitude scatter in the
lag.  NGC~5548 seems to lie close to the line in Fig.\@ 1. Scaling
the IR lag of Fairall 9 to NGC~5548 by $L^{1/2}$ would give a lag of
$\sim 200$ d. Since the grains are at $\sim 80$ ld (or less),
equation (3) of Barvainis (1987) shows that they would be at a
temperature of $\sim 2500$ K.  This is too high for grain survival.
In BLR shielding model, however, the continuum seen by the dusty
torus in AGNs is absorbed by the BLR, and the dust grains can be
much closer to the central source than they could otherwise be.  For
a factor of 2.7 reduction in the incident energy this allows the
dust to be at $\sim 120$ ld in NGC~5548. This is still greater than
our estimated 80 ld. While this is within the observational
uncertainties for NGC~5548, there are AGNs in Fig.\@ 1 with
substantially smaller dust lags for their luminosities, so an
additional effect is needed. As Kishimoto et al.\@ (2007) discuss,
dust can survive closer to the central source if the dust grains are
somewhat than those Barvainis (1987, 1992) considered. For larger
dust grains the IR emissivity rises substantially (see Draine \& Lee
1984) so they can cool much more efficiently and have lower
equilibrium temperatures. There is good evidence from X-ray optical
observations (Maiolino et al.\@ 2001a,b) and optical-UV extinctions
curves (Gaskell et al.\@ 2005; Gaskell \& Benker 2007) for dust
grains in AGNs being larger than those in the local interstellar
medium.

\section{CONCLUSIONS}

We have used simultaneous observations and the results of
multi-wavelength variability studies to consider in detail the most
probable SED of NGC~5548 during the highest state in the HST
monitoring in 1993.

We have argued that when the continuum of NGC~5548 (and other AGNs)
is corrected for Galactic and internal reddening, detailed
photoionization modelling shows that the strengths of all the major
emission lines can be explained by photoionization by the inferred
AGN continuum so long as the covering factors are large ($\sim
40$\%). The long-standing ``energy-budget problem'' can now be seen
to have been a consequence of not appreciating the significant
reddening of the continua of most AGNs and erroneously assuming a
spherical geometry. Because the emission-line strengths can be
explained by photoionization from the observed (or extrapolated)
continuum, there is no need for the additional heating source
suggested by Collin-Souffrin (1986), Joly (1987), and Dumont et
al.\@ (1998).

To reconcile our high covering factor with the lack of absorption
along the observer's line of sight we need a picture of the BLR
which is different from the widely assumed quasi-spherical
distribution of BLR clouds. Instead most of the BLR clouds
(especially those far from the AGN) have a flattened distribution
that extends out to the inner edge of the dusty torus. We have shown
that such a model is consistent with the observed line-continuum
transfer functions for NGC~5548.

In the geometry needed to explain the observed line strengths, the
cloud covering fraction is close to 100\% in the equatorial plane.
We have therefore proposed a shielding model of the BLR where the
inner clouds absorb much of the radiation before it reaches the
outer clouds or the torus.  This solves the problem of LOC models
not explaining the strong change in cloud conditions with radius.
The shielding BLR picture quantitatively reproduces the
ionization-dependent lags seen in NGC~5548.

The absorption by BLR gas also solves several problems with the
torus.  Allowance for BLR absorption increases the torus covering
factor calculated from energetics, and makes it slightly greater
than the mean BLR covering factor.  This reconciles the torus
covering factor with the fraction of obscured (type-2) AGNs and
explains why low-ionization by the BLR is never seen.  The reduced
flux seen by the torus also allows the dust to exist at smaller
radii and reduces the discrepancy between the calculated dust
sublimation radii and the radii of the inner edge of the torus found
from reverberation mapping.

Since the line ratios, equivalent widths, and the continuum shape of
NGC~5548 are similar to other AGNs (see, for example, Sergeev et
al.\@ 1999) we expect that these results can be generalized to most
other AGNs.

\acknowledgments

We are grateful to Ski Antonucci for many fruitful discussions of
reddening and torus covering factor issues. We are also grateful to
Bruce Draine and Joe Weingartner for helpful discussions of dust
issues, and to Bruce Draine for providing extinction models. We
thank Phil Uttley for providing the RXTE X-ray light curve and
Rebecca Harbison for assistance in collating optical photometry.
This research has been supported by the Space Telescope Science
Institute through grant AR-09926.01, and by the National Science
Foundation through grant AST 03-07912.


\begin{thebibliography}{}

\bibitem[Arav et al.\@(2003)]{arav03} Arav, N., Kaastra, J.,
Steenbrugge, K., Brinkman, B., Edelson, R., Korista, K.~T., \& de
Kool, M.\ 2003, \apj, 590, 174

\bibitem[Abramowicz et al.\@(1998)]{abramowicz98} Abramowicz, M.~A.,
Igumenshchev, I.~V., \& Lasota, J.-P.\ 1998, \mnras, 293, 443

\bibitem[Barvainis(1987)]{barvainis87}Barvainis, R.\ 1987, \apj,
320, 537

\bibitem[Barvainis(1992)]{barvainis92} Barvainis, R.\ 1992, \apj,
400, 502

\bibitem[Binette et al.\@(1993)]{binette93} Binette, L., Fosbury,
R.~A., \& Parker, D.\ 1993, \pasp, 105, 1150

\bibitem[Binette et al.\@(2005)]{binette05} Binette, L., Magris C.,
G., Krongold, Y., Morisset, C., Haro-Corzo, S., de Diego, J.~A.,
Mutschke, H., \& Andersen, A.~C.\ 2005, \apj, 631, 661

\bibitem[Bohlin et al.\@(1978)]{bohlin78} Bohlin, R.~C., Savage,
B.~D., \& Drake, J.~F.\ 1978, \apj, 224, 132

\bibitem[Boller et al.\@(1997)]{boller97} Boller, T., Brandt,
W.~N., Fabian, A.~C., \& Fink, H.~H.\ 1997, \mnras, 289, 393

\bibitem[Boller et al.\@(1996)]{boller96} Boller, T., Brandt,
W.~N., \& Fink, H.\ 1996, \aap, 305, 53

\bibitem[Bottorff \& Ferland(2002)]{bottorff02} Bottorff, M., \&
Ferland, G.\ 2002, \apj, 568, 581

\bibitem[Bottorff et al.\@(2000)]{bottorff00} Bottorff, M., Ferland,
G., Baldwin, J., \& Korista, K.\ 2000, \apj, 542, 644

\bibitem[Bottorff et al.\@(2002)]{bottorff02} Bottorff, M.~C.,
Baldwin, J.~A., Ferland, G.~J., Ferguson, J.~W., \& Korista, K.~T.\
2002, \apj, 581, 932

\bibitem[Brotherton et al.\@(2002)]{brotherton02} Brotherton, M.~S.,
Green, R.~F., Kriss, G.~A., Oegerle, W., Kaiser, M.~E., Zheng, W.,
\& Hutchings, J.~B.\ 2002, \apj, 565, 800

\bibitem[Cackett \& Horne(2006)]{cackett06} Cackett, E.~M., \&
Horne, K.\ 2006, \mnras, 365, 1180

\bibitem[Chiang \& Blaes(2003)]{chiang_blaes03} Chiang, J., \& Blaes,
O.\ 2003, \apj, 586, 97

\bibitem[Chiang et al.\@(2000)]{chiang00} Chiang, J., Reynolds,
C.~S., Blaes, O.~M., Nowak, M.~A., Murray, N., Madejski, G.,
Marshall, H.~L., \& Magdziarz, P.\ 2000, \apj, 528, 292

\bibitem[Choloniewski(1981)]{choloniewski81} Choloniewski, J. 1981, Acta Astron, 31, 293

\bibitem[Clavel et al.\@(1992)]{clavel92} Clavel, J., et al.\@\
1992, \apj, 393, 113

\bibitem[Collin-Souffrin(1986)]{collin-souffrin86} Collin-Souffrin, S.\
1986, \aap, 166, 115

\bibitem[Collin-Souffrin \& Lasota(1988)]{collin-souffrin_Lasota88}
Collin-Souffrin, S., \& Lasota, J.-P.\ 1988, \pasp, 100, 1041

\bibitem[Congiu et al.\@(2005)]{congiu05} Congiu, E., Geminale,
A., Barbaro, G., \& Mazzei, P.\ 2005, J. Phys. Conf. Ser., 6, 161

\bibitem[Crenshaw \& Blackwell(1990)]{crenshaw_blackwell90} Crenshaw, D.~M.,
\& Blackwell, J.~H., Jr.\ 1990, \apjl, 358, L37

\bibitem[Crenshaw \& Kraemer(2001)]{crenshaw01} Crenshaw, D.~M.,
\& Kraemer, S.~B.\ 2001, \apjl, 562, L29

\bibitem[Crenshaw et al.\@(2002)]{crenshaw02} Crenshaw, D.~M., et al.\@
\ 2002, \apj, 566, 187

\bibitem[Czerny et al.\@(2004)]{czerny04} Czerny, B., Li, J.,
Loska, Z., \& Szczerba, R.\ 2004, \mnras, 348, L54

\bibitem[Davidson(1972)]{davidson72} Davidson, K.\ 1972, \apj,
171, 213

\bibitem[De Zotti \& Gaskell(1985)]{dezotti85} de Zotti, G., \&
Gaskell, C.~M.\ 1985, \aap, 147, 1

\bibitem[Dietrich et al.\@(2001)]{dietrich01} Dietrich, M., et al.\@\
2001, \aap, 371, 79

\bibitem[Done \& Krolik(1996)]{done96} Done, C., \& Krolik,
J.~H.\ 1996, \apj, 463, 144

\bibitem[Done et al.\@(1995)]{done95} Done, C., Pounds, K.~A.,
Nandra, K., \& Fabian, A.~C.\ 1995, \mnras, 275, 417

\bibitem[Draine(2003a)]{draine03a} Draine, B.~T.\ 2003a, \apj, 598,
1017

\bibitem[Draine(2003b)]{draine03b} Draine, B.~T.\ 2003a, \apj, 598,
1026

\bibitem[Draine \& Lee(1984)]{1984ApJ...285...89D} Draine, B.~T., \& Lee,
H.~M.\ 1984, \apj, 285, 89

\bibitem[Dumont et al.\@(1998)]{dumont98} Dumont, A.-M.,
Collin-Souffrin, S., \& Nazarova, L.\ 1998, \aap, 331, 11

\bibitem[Edelson \& Malkan(1987)]{edelson87} Edelson, R.~A., \&
Malkan, M.~A.\ 1987, \apj, 323, 516

\bibitem[Ferland et al.\@(1998)]{ferland98} Ferland, G.~J.,
Korista, K.~T., Verner, D.~A., Ferguson, J.~W., Kingdon, J.~B., \&
Verner, E.~M.\ 1998, \pasp, 110, 761

\bibitem[Ferland \& Mushotzky(1982)]{ferland_mushotzky82} Ferland, G.~J.,
\& Mushotzky, R.~F.\ 1982, \apj, 262, 564

\bibitem[Ferland et al.\@(1992)]{ferland92} Ferland, G.~J.,
Peterson, B.~M., Horne, K., Welsh, W.~F., \& Nahar, S.~N.\ 1992,
\apj, 387, 95

\bibitem[Gaskell(1982)]{gaskell82} Gaskell, C.~M.\ 1982, \apj,
263, 79

\bibitem[Gaskell(1987)]{gaskell87} Gaskell, C.~M.\ 1987, in Astrophysical
Jets and their Engines, ed.\@ W. Kundt,  (Dordrecht: Reidel), p. 29

\bibitem[Gaskell(1988)]{gaskell88} Gaskell, C.~M.\ 1988, \apj,
325, 114

\bibitem[Gaskell(2000)]{gaskell00} Gaskell, C.~M.\ 2000, New
Astronomy Review, 44, 563

\bibitem[Gaskell(2007)]{gaskell07} Gaskell, C.~M.\ 2007, in The Central Engine of Active Galactic
Nuclei, ed.\@ L. C. Ho and J.-M. Wang (San Francisco: Astron. Soc.
Pacific), Vol. 373, p. 596

\bibitem[Gaskell \& Benker(2007)]{gaskell_benker07} Gaskell, C. M. \& Benker, A. J. 2007,
\apj, submitted [astro-ph/0711.1013]

\bibitem[Gaskell_et al.\@(2007)]{gaskell_et07} Gaskell, C. M., Goosmann, R. W., Merkulova, N. I.,
Shakhovskoy, N. M., \& Shoji, M. 2007, \apjl~Letts., ~submitted
[astro-ph/0711.1019]

\bibitem[Gaskell \& Klimek(2003)]{gaskell03} Gaskell, C.~M., \&
Klimek, E.~S.\ 2003, Astron. Astrophys. Trans., 22, 661

\bibitem[Gaskell \& Peterson(1987)]{gaskell87} Gaskell, C.~M., \&
Peterson, B.~M.\ 1987, \apjs, 65, 1

\bibitem[Gaskell \& Snedden(1997)]{gaskell_snedden97} Gaskell, C.~M., \&
Snedden, S.~A.\ 1997, BAAS, 29, 1252

\bibitem[Gaskell \& Sparke(1986)]{gaskell86} Gaskell, C.~M., \&
Sparke, L.~S.\ 1986, \apj, 305, 175

\bibitem[Gierli{\'n}ski \& Done(2004)]{gierlinski04} Gierli{\'n}ski,
M., \& Done, C.\ 2004, \mnras, 349, L7

\bibitem[2004]{glass04} Glass, I.~S.\ 2004, \mnras, 350, 1049

\bibitem[Goad \& Koratkar(1998)]{goad_koratkar98} Goad, M., \&
Koratkar, A.\ 1998, \apj, 495, 718

\bibitem[Goosmann \& Gaskell(2007)]{goosmann_gaskell07} Goosmann, R.~W.,
\& Gaskell, C.~M.\ 2007, \aap, 465, 129

\bibitem[Green et al.\@(1980)]{green80} Green, R. F., Pier, J. R., Schmidt, M., Estabrook, F. B.,
Lane, A. L., \& Wahlquist, H. D. 1980, ApJ, 239, 483

\bibitem[Haba et al.\@(2003)]{haba03} Haba, Y., Kunieda, H.,
Misaki, K., Terashima, Y., Kaastra, J.~S., Mewe, R., Fabian, A.~C.,
\& Iwasawa, K.\ 2003, \apj, 599, 949

\bibitem[Hamann \& Ferland(1999)]{hamann_ferland99} Hamann, F., \&
Ferland, G.\ 1999, \araa, 37, 487

\bibitem[Holt et al.\@(1980)]{holt80} Holt, S.~S., Mushotzky,
R.~F., Boldt, E.~A., Serlemitsos, P.~J., Becker, R.~H., Szymkowiak,
A.~E., \& White, N.~E.\ 1980, \apjl, 241, L13

\bibitem[Horne et al.\@(1991)]{horne91} Horne, K., Welsh, W.~F.,
\& Peterson, B.~M.\ 1991, \apjl, 367, L5

\bibitem[Joly(1987)]{joly87}Joly M., 1987, \aap, 184, 33

\bibitem[Kaastra \& Barr(1989)]{kaastra89} Kaastra, J.~S., \&
Barr, P.\ 1989, \aap, 226, 59

\bibitem[Kaspi \& Netzer(1999)]{kaspi_netzer99} Kaspi, S., \& Netzer,
H.\ 1999, \apj, 524, 71

\bibitem[Kishimoto et al.\@(2007)]{kishimoto07} Kishimoto, M., Honig,
S., Beckert, T., \& Weigelt, 2007, \aap submitted

\bibitem[Kleinmann \& Low(1970)]{kleinmann70} Kleinmann, D.~E., \&
Low, F.~J.\ 1970, \apjl, 159, L165

\bibitem[Koratkar \& Gaskell(1989)]{koratkar_gaskell89} Koratkar, A.~P.,
\& Gaskell, C.~M.\ 1989, \apj, 345, 637

\bibitem[Koratkar \& Gaskell(1991)]{koratkar_gaskell91a} Koratkar, A.~P.,
\& Gaskell, C.~M.\ 1991, \apjs, 75, 719

\bibitem[Koratkar \& Gaskell(1991b)]{koratkar_gaskell91b} Koratkar, A.~P.,
\& Gaskell, C.~M.\ 1991, \apj, 375, 85

\bibitem[Korista et al.\@(1995)]{korista95} Korista, K.~T., et al.\@\
1995, \apjs, 97, 285

\bibitem[Korista et al.\@(1997)]{korista97} Korista, K., Ferland,
G., \& Baldwin, J.\ 1997, \apj, 487, 555

\bibitem[Korista et al.\@(1998)]{korista98} Korista, K. T., Baldwin, J. A.,
\& Ferland, G. J. 1998, \apj, 507, 24

\bibitem[Korista \& Goad(2000)]{korista00} Korista, K.~T., \&
Goad, M.~R.\ 2000, \apj, 536, 284

\bibitem[Laor(997)]{laor97} Laor, A., 477, 93

\bibitem[Laor \& Draine(1993)]{laor_draine93} Laor, A., \& Draine,
B.~T.\ 1993, \apj, 402, 441

\bibitem[Laor(2007)]{laor07}Laor, A. 2007, in ``The Central Engine of Active Galactic Nuclei'', ed.\@ L. C. Ho
and J.-M. Wang, ASP Conf. Ser. 373, 000 [astro-ph/0702577]

\bibitem[Li \& Draine(2001)]{li_draine01} Li, A., \& Draine, B.~T.\
2001, \apjl, 550, L213

\bibitem[Lyutyi \& Doroshenko(1993)]{lyutyi93} Lyutyi, V.~M., \&
Doroshenko, V.~T.\ 1993, Astron. Lett., 19, 405


\bibitem[MacAlpine(1972)]{macalpine72} MacAlpine, G.~M.\ 1972,
\apj, 175, 11

\bibitem[MacAlpine(1981)]{macalpine81} MacAlpine, G.~M.\ 1981,
\apj, 251, 465

\bibitem[Magdziarz et al.\@(1998)]{magdziarz98} Magdziarz, P., Blaes,
O.~M., Zdziarski, A.~A., Johnson, W.~N., \& Smith, D.~A.\ 1998,
\mnras, 301, 179

\bibitem[Maiolino et al.\@(2001a)]{maiolino01b} Maiolino, R., Marconi, A., Salvati, M., Risaliti, G., Severgnini,
P., Oliva, E., La Franca, F., \& Vanzi, L. 2001a, \aap, 365, 28

\bibitem[Maiolino et al.\@(2001b)]{maiolino01b} Maiolino, R., Marconi, A., \& Oliva, E. 2001a, \aap, 365,
37

\bibitem[Maiolino et al.\@(2001c)]{maiolino01c} Maiolino, R., Salvati,
M., Marconi, A., \& Antonucci, R.~R.~J.\ 2001c, \aap, 375, 25

\bibitem[Maiolino et al.\@(2007)]{maiolino07} Maiolino, R., Shemmer,
O., Imanishi, M., Netzer, H., Oliva, E., Lutz, D., \& Sturm, E.\
2007, \aap, 468, 979

\bibitem[Markowitz et al.\@(2003)]{markowitz03} Markowitz, A.,
Edelson, R., \& Vaughan, S.\ 2003, \apj, 598, 935

\bibitem[Marshall et al.\@(1997)]{marshall97} Marshall, H.~L., et al.\@\ 1997, \apj, 479, 222

\bibitem[Marshall et al.\@(1995)]{marshall95} Marshall, H.~L.,
Fruscione, A., \& Carone, T.~E.\ 1995, \apj, 439, 90

\bibitem[Mathews \& Ferland(1987)]{mathews_ferland87} Mathews, W.~G., \&
Ferland, G.~J.\ 1987, \apj, 323, 456

\bibitem[McAlary et al.\@(1979)]{mcalary79} McAlary, C.~W.,
McLaren, R.~A., \& Crabtree, D.~R.\ 1979, \apj, 234, 471

\bibitem[McAlary et al.\@(1983)]{mcalary83} McAlary, C.~W.,
McLaren, R.~A., McGonegal, R.~J., \& Maza, J.\ 1983, \apjs, 52, 341

\bibitem[McKee \& Petrosian(1974)]{mckee74} McKee, C.~F., \&
Petrosian, V.\ 1974, \apj, 189, 17

\bibitem[Minezaki et al.\@(2006)]{minezaki06} Minezaki, T., Yoshii,
Y., Aoki, T., Kobayashi, Y., Suganuma, M., Enya, K., Tomita, H., \&
Peterson, B.~A.\ 2006, in AGN Variability from X-Rays to Radio
Waves, ed.\@ C. M. Gaskell, I. M. McHardy, B. M. Peterson and S. G.
Sergeev, ASP Conf. Ser., 360, 79

\bibitem[Moshir et al.\@(1990)]{moshir90} Moshir, M., et al.\@\
1990, \baas, 22, 1325

\bibitem[Mould et al.\@(2000)]{mould00} Mould, J.~R., et al.\@\
2000, \apj, 529, 786

\bibitem[Murphy et al.\@(1996)]{murphyu96} Murphy, E.~M., Lockman,
F.~J., Laor, A., \& Elvis, M.\ 1996, \apjs, 105, 369

\bibitem[Nazarova et al.\@(2004)]{nazarova04} Nazarova, L.~S.,
Bochkarev, N.~G., \& Gaskell, C.~M.\ 2004, Astron. \& Astrophys.
Trans., 23, 343

\bibitem[Nazarova et al.\@(2006)]{nazarova06} Nazarova, L.~S.,
Bochkarev, N.~G., Gaskell, C.~M., \& Klimek, E.~S.\ 2006, in AGN
Variability from X-Rays to Radio Waves, ed.\@ C. M. Gaskell, I. M.
McHardy, B. M. Peterson and S. G. Sergeev, ASP Conf. Ser., 360, 255

\bibitem[Netzer(1985)]{netzer85} Netzer, H.\ 1985, \apj, 289, 451

\bibitem[Netzer \& Laor(1993)]{netzer93} Netzer, H., \& Laor,
A.\ 1993, \apjl, 404, L51

\bibitem[Neugebauer et al.\@(1989)]{neugebauer89} Neugebauer, G.,
Soifer, B.~T., Matthews, K., \& Elias, J.~H.\ 1989, \aj, 97, 957

\bibitem[O'Brien et al.\@(1998)]{obrien98} O'Brien, P.~T., et al.\@\
1998, \apj, 509, 163

\bibitem[Oke(1970)]{Oke70} Oke, J.~B.\ 1970, \apjl, 161, L17

\bibitem[Oke(1974)]{Oke74} Oke, J.~B.\ 1974, \apjl, 189, L47

\bibitem[Oknyanskij et al.\@(1999)]{oknyanskij99} Oknyanskij, V.~L.,
Lyuty, V.~M., Taranova, O.~G., \& Shenavrin, V.~I.\ 1999, Astron.
Lett., 25, 483

\bibitem[Osmer(1979)]{osmer79} Osmer, P.~S.\ 1979, \apj, 227,18

\bibitem[Osterbrock(1977)]{osterbrock77} Osterbrock, D.~E.\ 1977,
\apj, 215, 733

\bibitem[Osterbrock(1989)]{osterbrock89} Osterbrock, D. E. 1989, Astrophysics of Gaseous Nebulae
and Active Galactic Nuclei, (Mill Valley: University Science Books),
p. 150

\bibitem[Osterbrock \& Shuder(1982)]{osterbrock_shuder82} Osterbrock,
D.~E., \& Shuder, J.~M.\ 1982, \apjs, 49, 149

\bibitem[Papadakis et al.\@(2002)]{papadakis02} Papadakis, I.~E.,
Petrucci, P.~O., Maraschi, L., McHardy, I.~M., Uttley, P., \&
Haardt, F.\ 2002, \apj, 573, 92

\bibitem[Peterson(2006)]{peterson06} Peterson, B.~M.\ 2006,
in AGN Variability from X-Rays to Radio Waves, ed.\@ C. M. Gaskell,
I. M. McHardy, B. M. Peterson and S. G. Sergeev, ASP Conf. Ser.,
360, 191

\bibitem[Peterson \& Wandel(1999)]{1999ApJ...521L..95P} Peterson, B.~M., \&
Wandel, A.\ 1999, \apjl, 521, L95

\bibitem[Pounds et al.\@(2003)]{pounds03} Pounds, K.~A., Reeves,
J.~N., Page, K.~L., Edelson, R., Matt, G., \& Perola, G.~C.\ 2003,
\mnras, 341, 953

\bibitem[Preite-Martinez \& Pottasch(1983)]{preite-martinez83}
Preite-Martinez, A., \& Pottasch, S.~R.\ 1983, \aap, 126, 31

\bibitem[Pringle \& Rees(1972)]{pringle_rees72} Pringle, J.~E., \&
Rees, M.~J.\ 1972, \aap, 21, 1

\bibitem[Reichert et al.\@(1994)]{reichert94} Reichert, G.~A., et al.\@\ 1994, \apj, 425, 582

\bibitem[Reimers et al.\@(2005)]{reimers05} Reimers, D., Hagen,
H.-J., Schramm, J., Kriss, G.~A., \& Shull, J.~M.\ 2005, \aap, 436,
465

\bibitem[Rieke(1978)]{rieke78} Rieke, G.~H.\ 1978, \apj, 226,
550

\bibitem[Rieke \& Low(1972)]{rieke72} Rieke, G.~H., \& Low,
F.~J.\ 1972, \apjl, 176, L95

\bibitem[Rodriguez-Pascual et al.\@(1997)]{rodrigues-pascual97}
Rodriguez-Pascual, P.~M., et al.\@\ 1997, \apjs, 110, 9

\bibitem[Rokaki et al.\@(1993)]{rokaki93} Rokaki, E., Collin-Souffrin, S., \& Magnan, C.\ 1993, \aap, 272, 8

\bibitem[Romanishin et al.\@(1995)]{romanishin95} Romanishin, W., et al.\@\ 1995, \apj, 455, 516

\bibitem[Schlegel et al.\@(1998)]{schlegel98} Schlegel, D.~J.,
Finkbeiner, D.~P., \& Davis, M.\ 1998, \apj, 500, 525

\bibitem[Schmitt et al.\@(1997)]{schmitt97} Schmitt, H.~R., Kinney,
A.~L., Calzetti, D., \& Storchi Bergmann, T.\ 1997, \aj, 114, 592

\bibitem[Sergeev et al.\@(1999)]{sergeev99} Sergeev, S.~G., Pronik,
V.~I., Sergeeva, E.~A., \& Malkov, Y.~F.\ 1999, \aj, 118, 2658

\bibitem[Shakura \& Syunyaev(1973)]{shakura_sunyaev73} Shakura, N.~I., \&
Syunyaev, R.~A.\ 1973, \aap, 24, 337

\bibitem[Shang et al.\@(2005)]{shang05} Shang, Z., et al.\@\ 2005,
\apj, 619, 41

\bibitem[Shapovalova et al.\@(2004)]{shapovalova04} Shapovalova, A.~I.,
et al.\@\ 2004, \aap, 422, 925

\bibitem[Shields(1973)]{shields73} Shields, G.~A.\ 1973,
Ph.D.~Thesis, Caltech

\bibitem[Shields et al.\@(1972)]{shields72} Shields, G.~A., Oke,
J.~B., \& Sargent, W.~L.~W.\ 1972, \apj, 176, 75

\bibitem[Shields \& Ferland(1993)]{shields93} Shields, J.~C., \&
Ferland, G.~J.\ 1993, \apj, 402, 425

\bibitem[Shuder(1982)]{shuder82} Shuder, J.~M.\ 1982, \apj, 259,
48

\bibitem[Smith et al.\@(1981)]{smith81} Smith, M.~G., et al.\@\
1981, \mnras, 195, 437

\bibitem[Snedden \& Gaskell(2004)]{snedden_gaskell04} Snedden, S., \&
Gaskell, C.\ 2004, in AGN Physics with the Sloan Digital Sky Survey,
ed.\@ G. T. Richards \& P. B. Hall, ASP Conf. Ser., 311, 197

\bibitem[Snedden \& Gaskell(2007a)]{snedden07a} Snedden, S. A. \& Gaskell, C. M. 2007, 668, 000
[astro-ph/0403174]

\bibitem[Snedden \& Gaskell(2007b)]{snedden07b} Snedden, S. A. \& Gaskell,C. M. 2007, submitted

\bibitem[Steenbrugge et al.\@(2003)]{steenbrugge03} Steenbrugge, K.~C.,
Kaastra, J.~S., de Vries, C.~P., \& Edelson, R.\ 2003, \aap, 402,
477

\bibitem[Suganuma et al.\@(2004)]{suganuma04} Suganuma, M., et al.\@\
2004, \apjl, 612, L113

\bibitem[Suganuma et al.\@(2006)]{suganuma06} Suganuma, M., et al.\@\
2006, \apj, 639, 46

\bibitem[Stoy(1933)]{Stoy33} Stoy, R.~H.\ 1933, \mnras, 93,
588

\bibitem[Uttley et al.\@(2003)]{uttley03} Uttley, P., Edelson, R.,
McHardy, I.~M., Peterson, B.~M., \& Markowitz, A.\ 2003, \apjl,584,
L53


\bibitem[Walter \& Courvoisier(1990)]{walter_corvoisier90} Walter, R., \&
Courvoisier, T.~J.-L.\ 1990, \aap, 233, 40

\bibitem[Walter et al.\@(1994)]{walter94} Walter, R., Orr, A.,
Courvoisier, T.~J.-L., Fink, H.~H., Makino, F., Otani, C., \&
Wamsteker, W.\ 1994, \aap, 285, 119

\bibitem[Wanders et al.\@(1997)]{wanders97} Wanders, I., et al.\@\
1997, \apjs, 113, 69

\bibitem[Wanders \& Peterson(1996)]{wanders96} Wanders, I., \&
Peterson, B.~M.\ 1996, \apj, 466, 174

\bibitem[Welsh et al.\@(2007)]{welsh07} Welsh, W.~F., Martino,
D.~L., Kawaguchi, G., \& Kollatschny, W.\ 2007, to appear in The
Central Engine of AGNs, ed.\@ J.-M. Wang, ASP Conf. Ser.
[astro-ph/0707.2606]

\bibitem[Weingartner \& Draine(2001)]{weingartner01} Weingartner,
J.~C., \& Draine, B.~T.\ 2001, \apj, 548, 296

\bibitem[Winkler et al.\@(1992)]{winkler92} Winkler, H., Glass,
I.~S., van Wyk, F., Marang, F., Jones, J.~H.~S., Buckley, D.~A.~H.,
\& Sekiguchi, K.\ 1992, \mnras, 257, 659

\bibitem[Zanstra(1931)]{zanstra31} Zanstra, H.\ 1931, Zeitschrift
fur Astrophysik, 2, 1

\bibitem[Zheng et al.\@(1997)]{zheng97} Zheng, W., Kriss, G.~A.,
Telfer, R.~C., Grimes, J.~P., \& Davidsen, A.~F.\ 1997, \apj, 475,
469

\end{thebibliography}
\end{document}